\documentclass[aps,prd,amsmath,eqsecnum,floatfix,onecolumn,nofootinbib,11pt]{revtex4-2}

\usepackage{amsmath,amssymb,mathrsfs,esvect,slashed}
\usepackage{calc,array,booktabs}
\usepackage{graphicx}
\usepackage{pifont}
\usepackage{setspace}
\usepackage{soul}
\usepackage{outlines}
\usepackage{latexsym}

\usepackage{subcaption}

\usepackage[dvipsnames]{xcolor}
\usepackage{tikz,tikz-3dplot}
\usetikzlibrary{arrows.meta}

\usepackage{listings}
\usepackage{tcolorbox}
\tcbuselibrary{listings}
\lstdefinelanguage{GuestFortran}{language=Fortran,morekeywords={implicit,none,integer,real,do,enddo,call,subroutine,return},sensitive=true}

\usepackage{hyperref}
\hypersetup{pdfpagemode=UseOutlines,bookmarksopen=true,pdfstartview=FitH,colorlinks=true,linkcolor=blue,citecolor=blue,urlcolor=blue}

\newcommand{\be}{\begin{eqnarray}}
\newcommand{\ee}{\end{eqnarray}}

\newcommand{\bea}{\begin{eqnarray}}
\newcommand{\eea}{\end{eqnarray}}

\newcommand{\beq}{\begin{equation}}
\newcommand{\eeq}{\end{equation}}

\newcommand{\norm}[1]{\left\lVert#1\right\rVert}

\newlength{\oldwidth}
\newlength{\oldheight}

\newcommand\ze{\zeta}

\begin{document}

\title{Geometrical and Optical Design of the GUEST Laser-Ranged Test Masses for the Detection of Gravitational Waves in the $\mu$Hz range}

\author{Dario Vetrano}
\email{dvetrano@lnf.infn.it}
\affiliation{
{\small Laboratori Nazionali di Frascati (LNF) - Istituto Nazionale di Fisica Nucleare (INFN), Via Enrico Fermi, 54, Frascati (RM) 00044, Italy}
}

\author{Giada Bargiacchi}
\email{gbargiacchi@lnf.infn.it}
\affiliation{
{\small Laboratori Nazionali di Frascati (LNF) - Istituto Nazionale di Fisica Nucleare (INFN), Via Enrico Fermi, 54, Frascati (RM) 00044, Italy}
}

\author{Nicol\`o Burzill\`a}
\email{burzilla@lnf.infn.it}
%\email{nicolo.burzilla@roma2.infn.it}
\email{nburzilla@outlook.it}
\affiliation{
{\small Laboratori Nazionali di Frascati (LNF) - Istituto Nazionale di Fisica Nucleare (INFN), Via Enrico Fermi, 54, Frascati (RM) 00044, Italy}
}

\author{Roberto Campagnola}
\email{Roberto.Campagnola@lnf.infn.it}
\affiliation{
{\small Laboratori Nazionali di Frascati (LNF) - Istituto Nazionale di Fisica Nucleare (INFN), Via Enrico Fermi, 54, Frascati (RM) 00044, Italy}
}

\author{Francesco Del Ferraro Iacobelli}
\email{francesco.delferraroiacobelli@lnf.infn.it}
\affiliation{
{\small Laboratori Nazionali di Frascati (LNF) - Istituto Nazionale di Fisica Nucleare (INFN), Via Enrico Fermi, 54, Frascati (RM) 00044, Italy}
}

%\author{Diego Blas Temi\~{n}o}
\author{Diego Blas}
\email{dblas@ifae.es}

\affiliation{
(GUEST Principal Investigator, PI)
}

\affiliation{
{\small Institut de F\'{i}sica d’Altes Energies (IFAE), The Barcelona Institute of Science and Technology, Campus UAB, Bellaterra (Barcelona) 08193, Spain}
}
\affiliation{
Instituci\'{o} Catalana de Recerca i Estudis Avan\c{c}ats (ICREA), Passeig Ll\'{u}ıs Companys 23, 08010 Barcelona, Spain
}

\author{Luca Porcelli}
\email{luca.porcelli@lnf.infn.it}
\affiliation{
{\small Laboratori Nazionali di Frascati (LNF) - Istituto Nazionale di Fisica Nucleare (INFN), Via Enrico Fermi, 54, Frascati (RM) 00044, Italy}
}

\author{Simone Dell'Agnello}
\email{simone.dellagnello@lnf.infn.it}
\affiliation{
{\small Laboratori Nazionali di Frascati (LNF) - Istituto Nazionale di Fisica Nucleare (INFN), Via Enrico Fermi, 54, Frascati (RM) 00044, Italy}
}
\date{\today}

%\center {To be submitted to Arxiv.}

%\begin{abstract}
%\quad \quad \quad \quad \quad \quad \quad \quad \quad \quad \quad \quad {\bf (To be submitted to Arxiv)} \\
\begin{abstract}
{\centering
\textbf{Abstract}\\[1.5em]
}
\par \noindent
% The GUEST (Gravitational Universe Exploration with Satellite Tracking) mission aims at detecting gravitational waves in the yet almost unexplored $\mu$Hz frequency range through high-precision orbital tracking. The unprecedented scientific objectives of this experiment demand equally advanced optical solutions: the highly eccentric orbit imposes dual measurement regimes; a low-velocity, long-range scenario at apogee and a high-velocity, short-range scenario at perigee. The substantial disparity between these conditions is addressed through a novel optical architecture incorporating two distinct types of Corner-Cube Retroreflectors (CCRs) with 1" and 2" apertures (CCR1 and CCR2, respectively).
% The GUEST (Gravitational Universe Exploration with Satellite Tracking) mission aims at detecting gravitational waves in the yet almost unexplored $\mu$Hz frequency range through high-precision orbital tracking. The unprecedented scientific objectives of this experiment demand equally advanced optical solutions: the highly eccentric orbit imposes dual measurement regimes; a low-velocity, long-range scenario at apogee and a high-velocity, short-range scenario at perigee. The substantial disparity between these conditions is addressed through a novel optical architecture incorporating two distinct types of Corner-Cube Retroreflectors (CCRs) with 1" and 2" apertures (CCR1 and CCR2, respectively).
The GUEST (Gravitational Universe Exploration with Satellite Tracking) mission aims at detecting gravitational waves in the yet almost unexplored $\mu$Hz frequency range through high-precision orbital tracking. The unprecedented scientific objectives of this experiment demand equally advanced optical solutions: the highly eccentric orbit imposes dual measurement regimes; a low-velocity, long-range scenario at apogee and a high-velocity, short-range scenario at perigee. The substantial disparity between these conditions is addressed through a novel optical architecture incorporating two distinct types of Corner-Cube Retroreflectors (CCRs) with 1" and 2" apertures (CCR1 and CCR2, respectively).
The proposed layout has been designed to ensure isotropy and homogeneity of the optical response while maintaining efficient packing over the spherical surface of the spacecraft. The resulting Optical Cross-Section (OCS) has been evaluated using a custom reduced-order model, which captures the behavior of individual CCRs under varying illumination conditions.
This model has been calibrated by reconstructing the optical response of the LARES-2 satellite and has enabled a detailed assessment of the contributions of each CCR type along the GUEST orbit. Finally, sensitivity analyses have been conducted to investigate the impact of key design parameters, including the ratio between CCR types, laser wavelength, spacecraft diameter, Dihedral Angle Offsets (DAOs), and mechanical configurations.
\end{abstract}

\maketitle

\section{Introduction}
\label{sec:introduction}

GUEST~\cite{GUEST_white_paper} is a passive-satellite laser-ranging mission proposal whose primary objective is the detection of gravitational waves (GWs) with strong synergies with dark-matter searches, tests of gravitation, and geodesy. Specifically, gravitational waves from mergers of binary supermassive black holes may become accessible for the first time in the largely unexplored  $\mu$Hz frequency range. In particular, GUEST is designed to open access to GWs from slowly evolving supermassive black-hole binaries at frequencies within the so called ``$\mu$Hz gap'', below the sensitivity range of space-based interferometers like LIGO-Virgo-KAGRA (LVK) at $\sim 10 \div 10^3$ Hz~\cite{LIGO2016} and above the current sensitivity of the Pulsar Timing Array (PTA) centered at nHz frequencies~\cite{NANOGrav2023}. Similarly, the future ESA flagship mission LISA will cover the mHz band~\cite{LISA2024}.
Furthermore, GUEST will be able to probe ultra-light dark-matter models in regions of parameter space that are currently unconstrained by any other experiment, while also advancing tests of General Relativity and searches for new particles with a technique complementary to particle accelerators~\cite{GUEST_white_paper,fosterBlas2025}.

The GUEST detection concept is not based on interferometry but rather on resonance effects of orbital dynamics; in this sense, the orbits themselves act as a detector. In particular, the gravitational wave acts as a weak time-dependent perturbation of the Earth--satellite gravitational system, while the orbit provides a dynamical response that can accumulate over many orbital cycles. Given $a$, the semi-major axis of the orbit, and considering a GW of frequency of the same order of magnitude of the orbital frequency we have that $\lambda_{\mathrm{GW}} \sim c \, T_{\mathrm{GW}} \gg v_0\, T_{\mathrm{orb}} \sim 2\pi a $. In this regime a weak GW acts on the Earth-satellite binary system as a perturbing acceleration of the form ~\cite{Hui2013,Blas2022,Blas2022a,Foster2025a,Foster2025b,GUEST_white_paper}
\begin{equation}
a^{i}
=
\frac{1}{2}\ddot{h}_{ij}(t) \, r^{j},
\end{equation}
where $r^{j}$ displacement vector between the orbiting bodies and $h_{ij}(t)$ is the transverse-traceless metric perturbation. A GW with frequency $\omega_{\mathrm{GW}}$ that is harmonic with the orbital frequency can resonantly couple to the orbital motion inducing a time evolution of the six orbital elements.
The resonance condition 
\begin{equation}
\omega_{\mathrm{GW}} \simeq \frac{2\pi n}{T_{\mathrm{orb}}},
\qquad n=1,2,3 \dots \;,
\end{equation}
allows for an efficient transfer of energy between the GW field and the binary system. These orbital perturbations accumulate over time, manifesting, in particular, as a quadratic drift of the true anomaly.
The basic physical picture of GUEST is therefore that of a gravitational analogue of a resonant receiving antenna, where the orbit responds to the incoming GW and stores its effect in the subsequent evolution of the orbital motion~\cite{Hui2013,Blas2022,Blas2022a,Foster2025a,Foster2025b,GUEST_white_paper}. 
GUEST will exploit this effect by deploying two passive satellites in highly eccentric orbits ($e \sim 0.75$) with periods of approximately $T \sim 33.8$ hours, corresponding to a frequency $\nu_{\mathrm{GW}}=\omega/2\pi\sim 1.6 \cdot 10^{-5}$ Hz for the $n=2$ harmonic. A sufficiently large orbital period is key to access the sensitivity of the ``$\mu$Hz gap'', while the high eccentricity allows to accelerate the quadratic growth of the orbital parameters, enhancing such effect. At the same time, high eccentricity excites higher harmonics, subject to the condition $n \ll 1/v_0$ required by the long-wavelength approximation.
~\cite{Blas2022,Blas2022a,Hui2013,Foster2025a,Foster2025b}. On the other hand, the deployment of two satellites and their relative orbital geometry helps decorrelate some of the systematics while providing more complete sky coverage.
The satellites will be tracked daily by Satellite Laser Ranging (SLR) stations of the International Laser Ranging Service (ILRS) over a $30$-year mission lifetime \cite{GUEST_white_paper}.
The resulting SLR data will enable a precise determination of the forces acting on the probes. 
Over this timescale, the ranging measurements provide a long time series from which the orbital evolution can be reconstructed with high precision, simultaneously accounting for the gravitational and non-gravitational forces required for precise orbit determination.  
The use of highly eccentric orbit, however, introduces a particularly challenging optical-ranging problem. The spacecraft velocity and Earth--satellite distance vary by large factors between perigee and apogee. At perigee, the high transverse velocity produces a large velocity-aberration angle of about $48$~$\mu$rad and strongly modifies the fraction of the CCR far-field diffraction pattern intercepted by the transmitting and receiving station. At apogee, in contrast, the velocity aberration is much smaller ($\sim7$ $\mu$rad) but the much larger distance ($\sim86700$ km) strongly reduces the received photon flux. A solution to overcome this challege, presented in~\cite{GUEST_white_paper}, was to employ two populations of CCRs with different apertures, providing a response adapted to the different ranging conditions encountered along the orbit.
The optical performance of the retroreflector system and their geometrical distribution is therefore an important component of the overall GUEST measurement chain. The ranging precision required to reconstruct the long-term orbital perturbations depends on the ability of the satellite to provide sufficiently strong and stable laser returns over the full orbit. In particular, the OCS must be evaluated as a function of the changing range, velocity aberration, illumination geometry, and CCR optical response. The present work focuses on this aspect of the GUEST mission and develops a reduced-order model to characterize the optical response of the proposed dual-CCR configuration.

The remainder of this paper is organized as follows. Section~\ref{sec:optical_requirements} introduces the optical requirements imposed by the GUEST orbital configuration and describes the geometrical arrangement of the two CCR populations on the spherical spacecraft. Section~\ref{sec:optical_performance} discusses the optical response of the CCRs, with particular emphasis on the Far-Field Diffraction Pattern (FFDP), velocity aberration, and incidence angle. Section~\ref{sec:reduced_order_model} introduces the reduced-order model used to estimate the spacecraft OCS and describes the parametrization of the incidence-angle and velocity-aberration effects, including the treatment of dihedral angle offsets (DAOs). In Section~\ref{sec:model_validation}, the model is validated and calibrated against the LARES-2 satellite. Section~\ref{sec:guest_baseline_configuration} presents the baseline GUEST configuration and the sensitivity studies performed by varying the CCR population ratio, laser wavelength, spacecraft size, and CCR DAOs. The mechanical configuration, including the use of protrusions for the CCR2 population, is discussed in Section~\ref{sec:mechanical_configuration}. Finally, Section~\ref{sec:conclusions} summarizes the main results and discusses future developments of the design phases.

\section{Satellite Optical Requirements}
\label{sec:optical_requirements}

The GUEST mission concept relies on two dense, passive, spherical satellites tracked by SLR in highly eccentric, long-period Earth orbits. This orbital regime imposes an optical-design problem that is not encountered in nearly circular geodetic satellites \cite{Degnan2023}: the velocity aberration of the returned laser beam varies significantly along the orbit. A single population of CCRs cannot simultaneously maximize the OCS at perigee, where the transverse velocity and aberration angle are largest, and at apogee, where the slant range is largest and photon return is intrinsically weak \cite{Degnan2023}. For this reason, the proposed solution for this unprecedented laser-ranging scenario consists of a dual CCR population mounted on the satellite, with two different front-face diameters.

\subsection{Construction of the Spherical-cap Model}
\label{sec:spherical_cap_model}

For practical modeling purposes, all of the satellite designs considered in this work are fixed to a spherical shape. Proceeding this way, models developed for previous satellite missions (like the LARES-2 mission) can be inherited in a straightforward way. 
The problem can therefore be modeled as searching for configurations of the two CCR populations on the spherical surface of the satellite such that the following two criteria are satisfied:
\begin{enumerate}
    \item Isotropy, as the OCS should not exhibit large fluctuations among different possible Lines of Sight (LoS).
    \item Efficient packing, as the spherical surface should accommodate as many CCRs as possible in order to maximize the OCS, while also accounting for mounting and mechanical constraints.
\end{enumerate}

Such a design can be formulated geometrically as the arrangement, on a spherical surface, of two populations of spherical caps with prescribed semi-apertures. The semi-aperture associated with each cap, which represents one CCR population, follows from the chord--angle relation
\begin{equation}
    D = 2R \sin\left(\frac{\delta}{2}\right),
    \label{eq:angle_chord_relation}
\end{equation}
where $D$ is the chord length on a sphere of radius $R$ and $\delta$ is the corresponding angular aperture.

\begin{figure}
    \centering
    \includegraphics[width=0.5\linewidth]{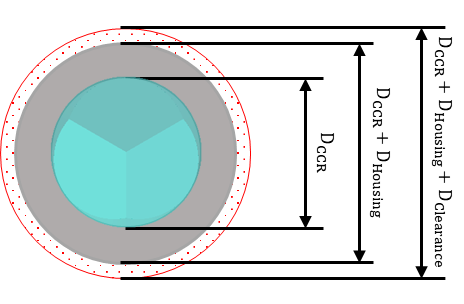}
    \caption{Schematic representation of the radii contributing to the diameter of each spherical cap, projected onto the plane of the CCR front face.}
    \label{fig:ccr_mount_clearance}
\end{figure}

For each CCR population, the effective cap radius is computed as the sum of the following contributions, as illustrated in Fig.~\ref{fig:ccr_mount_clearance}:
\begin{itemize}
    \item[\ding{226}] the CCR front-face radius: $D_{CCR}$;
    \item[\ding{226}] the radial thickness of the CCR housing: $D_{Housing}$;
    \item[\ding{226}] the required clearance: $D_{Clearance}$.
\end{itemize}

The CCR front-face diameters for each of the two populations are set in this work to $1\,\mathrm{in}$ and $2\,\mathrm{in}$, hereafter denoted as CCR1 and CCR2, respectively. These values were selected on the basis of INFN -LNF retroreflector heritage and preliminary qualification constraints. Mount thickness and clearance values were also derived from previous heritage \cite{ciufolini_lares2,arnold_lares_transfer_2015} and technical documentation of the mounting setup. The minimum-clearance condition is therefore reformulated as a tangency condition among spherical caps.

Since the arrangement is defined on a sphere, the geometry can be described equivalently in terms of chord lengths $D$ or aperture angles $\delta$, which are directly related through Eq.~\eqref{eq:angle_chord_relation}.

\subsection{Geometrical Arrangement of Two Spherical-cap Populations on a Sphere}
\label{sec:cap_arrangement}

In order to satisfy the packing and isotropy requirements efficiently, the first step is to construct a discretization of the sphere that does not introduce preferred directions. This endeavor of evenly distributing points on a sphere is of great interest in many fields of study and has been extensively explored~\cite{points_on_sphere}. The most intuitive choice, namely a grid of equally spaced points in azimuth and latitude coordinate space, trivially fails to satisfy this requirement because of the two well-known singularities at the poles. A better solution that is also commonly adopted in computer graphics consists in unfolding a Fibonacci spiral along a chosen axis~\cite{fibonacci_sphere, points_on_sphere}. Although this method is unaffected by the singularity issue encountered in the aforementioned spherical grid, it still falls short of the isotropy condition due to its dependence on the choice of a given axis. Solutions based on polyhedra should also be discarded because of the discrete nature of the available tessellations, which conflicts with the requirement of efficient packing. Indeed, unless particularly favourable size ratios occur, any feasible configuration with non-overlapping spherical caps typically overshoots the required first-neighbor clearance, thereby degrading the packing efficiency.

The proposed solution consists of two steps:
\begin{enumerate}
    \item Create a coarse grid of points following the Thomson distribution, that is an approximate solution to the Thomson problem of arranging electrons on an electrically conducting sphere~\cite{thomson_solutions,points_on_sphere};
    \item Identify, within this coarse grid, the fundamental spherical-triangular tiles and populate each tile with the same arbitrary number of points in a pattern defined relative to the tile vertices.
\end{enumerate}

\begin{figure}
    \centering
    \includegraphics[width=\linewidth]{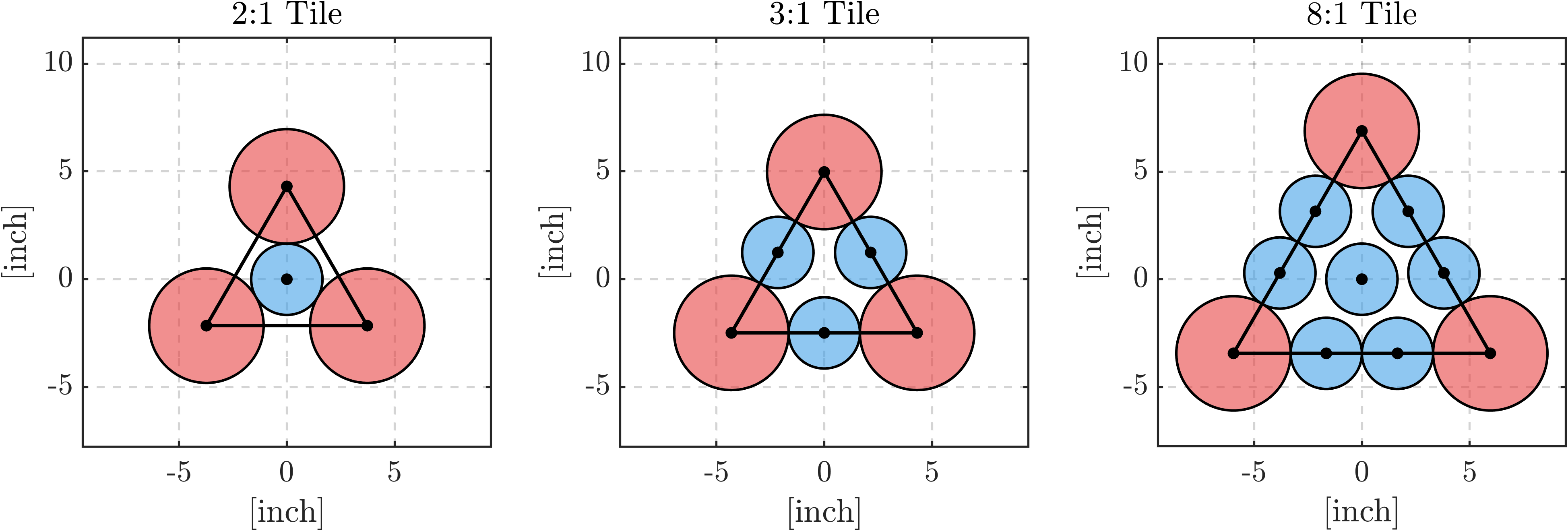}
    \caption{Representative tile configurations available for the satellite tessellation, corresponding to different CCR1/CCR2 population ratios.}
    \label{fig:different_ccr_ratios}
\end{figure}

This procedure makes it possible to construct, on the spherical surface triangulated according to the Thomson solution, configurations with different ratios between the CCR1 and CCR2 populations. The global ratio resulting from the complete tiling of the sphere can be inferred by summing the fractional contributions contained in each triangular tile and then taking their ratio, as illustrated in Fig.~\ref{fig:different_ccr_ratios}, where each circle at a vertex is shared between six adjacent tiles and thus must be counted as $1/6$. Accordingly, each circle on a side must be counted as $1/2$. The Thomson solutions used in this work were adopted from Wales and Ulker~\cite{thomson_solutions}.

\section{Corner-Cube Retroreflector Optical Response}
\label{sec:optical_performance}
\begin{figure}[htbp]
    \centering
    \begin{subfigure}[b]{0.35\textwidth}
        \centering
        \includegraphics[width=\textwidth, trim=0 0.5cm 0 0,clip]{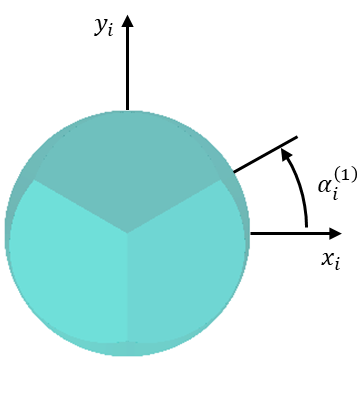}
        \caption{}
        \label{fig:ff_ccr_ref_sys}
    \end{subfigure}
    \hfill
    \begin{subfigure}[b]{0.35\textwidth}
        \centering
        \includegraphics[width=\textwidth,trim=1cm 2cm 0 0,clip]{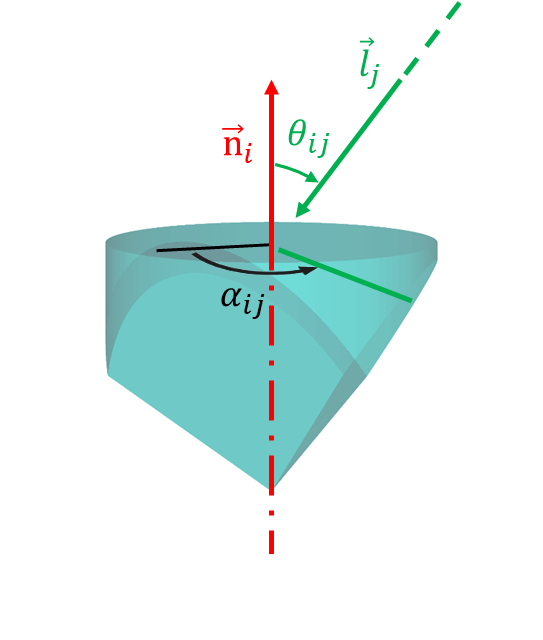}
        \caption{}
        \label{fig:laser_on_ccr}
    \end{subfigure}

    \caption{Schematic representation of a plane wave impinging on the front face of a CCR.}
    \label{fig:ref_sys_ccr}
\end{figure}

When a laser beam impinges on the front surface of a CCR, different portions of the wavefront travel along different optical paths inside the reflector before being returned through the same aperture.
This produces a non-trivial diffraction pattern~\cite{Degnan2023} which, under the observing conditions considered here, occurs in the far-field/Fraunhofer regime. Therefore, knowledge of the Far-Field Diffraction Pattern (FFDP) of a CCR provides information on the angular redistribution of the laser intensity when the beam is detected during its return path. The OCS, denoted by $\sigma$, is related to the laser intensity through the relation~\cite{Degnan2023, Goodman2005}:
\begin{equation}
    \sigma(\chi, \eta) = \frac{4\pi}{\lambda^2} \cdot \frac{I(\chi, \eta)}{\|E_0\|^2},
\end{equation}
where,
\begin{equation}
    I(\chi,\eta) = \norm{\iint_{\mathrm{aperture}} \vec{E}(x,y) e^{ik(\chi x + \eta y) } \operatorname{d}x \operatorname{d}y }^2.
\label{eq:diffraction_integral}
\end{equation}
Here $(x, y)$ is a point on the CCR aperture (see Fig. \ref{fig:ff_ccr_ref_sys}), $\vec{E}(x,y)$ is the outgoing complex electric field, ($\chi$, $\eta$) are far-field angular coordinates as defined in Fig. \ref{fig:ffdp_ocs_vs_va_panels}, $E_0$ is the amplitude of the laser impinging on the CCR\footnote{As we are considering only monochromatic, completely polarized waves, its definition is unambiguous.}, and $\lambda$ is the wavelength of the laser. The non-zero angular displacement between the incident and returned beam is caused by the relative motion between the CCR and the observer, i.e. the laser-ranging station, and is referred to as Velocity Aberration (VA)~\cite{Degnan2023}, denoted here by $\mu$. A useful approximation for the velocity aberration is
\begin{equation}
    \mu \simeq 2\,\frac{v_{\perp}}{c},
    \label{eq:velocity_aberration}
\end{equation}
where $v_{\perp}$ is the magnitude of the satellite velocity component perpendicular to the line connecting the laser-ranging station and the spacecraft, evaluated in a reference frame fixed to the station. This approximation assumes that $v_{\perp}$ remains approximately constant during the round trip of the laser pulse, which is generally valid.
Nonetheless, given our previous definition of $(\chi, \eta)$ as angular cartesian coordinates on the far-field, we see that we can define also angular polar coordinates on such plane as:
\begin{equation}
        \mu = \sqrt{\chi^2 + \eta^2}, \qquad 
        \varphi = \operatorname{atan2}(\eta, \chi).
\end{equation}
The orientation of a generic $i$-th CCR in space may be fixed through the normal vector to its front-face aperture $\vec n_i$ and the azimuth angle between the reference direction $x_i$ and the first edge of the CCR, $\alpha_i^{(1)}$ (Fig. \ref{fig:ff_ccr_ref_sys}). For a given LoS $\vec l_j$, the normal vector $\vec n_i$ may be misaligned with the LoS by two angles: the azimuth $\alpha_{ij}$ and the co-latitude $\theta_{ij}$, as shown in Fig.~\ref{fig:laser_on_ccr}.

Summarizing, the OCS of a CCR observed under such conditions is affected by three angles~\cite{Degnan2023}:
\begin{itemize}
    \item[\ding{226}] CCR relative clocking: upon variation of the angle $\alpha_i^{(1)} - \alpha_{ij}$, the FFDP rotates and deforms with a threefold periodicity per revolution; indeed rotations by $120^\circ$ leave the FFDP unchanged unless different Dihedral Angle Offsets (DAOs) are present, as we will see later.
    \item[\ding{226}] Laser incident angle, $\theta_{ij}$: the returned OCS strictly decreases as the incidence angle increases.
    \item[\ding{226}] Velocity aberration $\mu$: the return decreases with respect to the peak OCS at $\mu = 0\,\mathrm{rad}$.
\end{itemize}

\section{Reduced-order observing model}
\label{sec:reduced_order_model}

To reduce the complexity of the problem, FFDPs are usually averaged over the far-field azimuth~\cite{arnold_lares_transfer_2015} in order to obtain a curve of OCS as a function of $\mu$:

\begin{equation} \label{eq:definition_sigma_bar}
    \bar \sigma(\mu) = \frac{1}{2\pi} \int_0^{2\pi} \sigma(\mu,\varphi) \operatorname{d}\varphi.
\end{equation}

This is justified by the fact that, when a Laser Retroreflector Array (LRA) is observed, more than one CCR contributes to the return signal, and the individual CCRs generally have different azimuthal orientations. In practice, the azimuthal orientation of the front face is one of the parameters that can be exploited for OCS optimization~\cite{Degnan2023}; it is only partially disregarded at this preliminary stage of the design process.

\begin{figure}
    \centering
    \includegraphics[width=\linewidth]{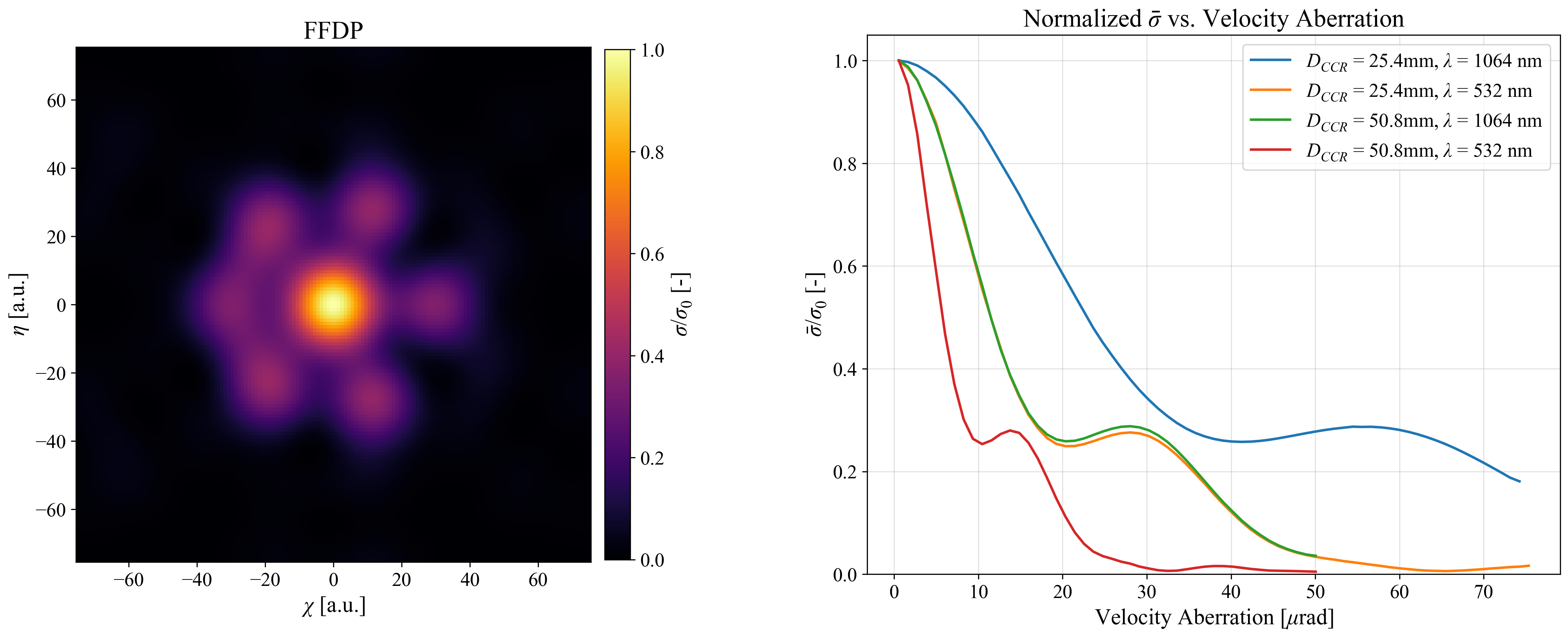}
    \caption{Normalized Far-field Diffraction Pattern (FFDP) of a CCR and normalized $\bar \sigma(\mu)$ curves as defined by Eq. \eqref{eq:definition_sigma_bar} for different combinations of $D_{CCR}$ and $\lambda$.}
    \label{fig:ffdp_ocs_vs_va_panels}
\end{figure}

Figure~\ref{fig:ffdp_ocs_vs_va_panels} shows normalized $\bar\sigma(\mu)$ curves associated with different FFDPs for the two CCR populations. These FFDPs were computed with an internal ray-tracing software that was validated against laboratory FFDP measurements of LARES-2 CCRs. The same software is also used in the following sections to quantify the effect of off-design observing conditions. The slight mismatch between the orange ($D_{CCR} = 25.4\,\mathrm{mm}, \lambda=532\,\mathrm{nm}$) and green ($D_{CCR} = 50.8\,\mathrm{mm}, \lambda=1064\,\mathrm{nm}$) curves is not expected from theory, and it is due to different resolutions in the simulations of the discretized CCR (a bigger CCR requires more rays, hence more computational time); nonetheless, the percentage difference has been kept always lower than 5\%. The peak values of OCS $\sigma_0$ for each combination of $D_{CCR}$ and $\lambda$ are reported in Table \ref{tab:peak_ocs_values}.

\begin{table}[htpb]
    \centering
    \begin{tabular*}{0.75\linewidth}{@{\extracolsep{\fill}} c c c @{}}
    \hline
    CCR Diameter [mm] & Laser wavelength [nm] & Peak OCS $[10^6 \,\mathrm{m^2}]$ \\
    $D_{CCR}$ & $\lambda$  & $\sigma_{0}$ \\
    \hline
         \qquad 25.4 & 532 & 2.77 
         \qquad  \\
         \qquad 50.8 & 532 & 45.6 
         \qquad   \\ 
         \qquad 25.4 & 1064 & 0.69
         \qquad   \\
         \qquad 50.8 & 1064 & 11.1
         \qquad   \\ \hline
    \end{tabular*}
    \caption{Peak values of OCS $\sigma_0$ for different CCR diameters and laser wavelengths.}
    \label{tab:peak_ocs_values}
\end{table}

For a given LoS, $\vec l_j$, the total spacecraft (S/C) OCS is approximated as
\begin{equation}
    \sigma_{\mathrm{S/C},j}
    =
    \sum_{i \in \mathcal{V}_j}
    \sigma_i(\theta_{ij},\mu),
    \label{eq:total_sc_ocs}
\end{equation}
where $\mathcal{V}_j$ is the set of CCRs visible from the $j$-th LoS\footnote{The dispersion of $\sigma_{\mathrm{S/C},j}$ over the set of LoS can be used as a measure of the satellite isotropy, as discussed in Sec.~\ref{sec:model_validation}.}. The contribution of the $i$-th CCR is modeled as
\begin{equation}
    \sigma_i(\theta_{ij},\mu)
    =
    k^{\mathrm{inc}}(\theta_{ij})\,
    k^{\mathrm{ab}}(\mu)\,
    \sigma_{0,i},
    \label{eq:single_ccr_ocs}
\end{equation}
with 
\begin{equation}
    0 \leq k^{\mathrm{inc}}(\theta_{ij}) \leq 1,
    \qquad
    0 \leq k^{\mathrm{ab}}(\mu) \leq 1.
    \label{eq:k_factor_bounds}
\end{equation}
weighting functions, representing the diminished optical return due to non normal incidence and velocity aberration, respectively. Eq.~\eqref{eq:single_ccr_ocs} represents the core assumption of our model, as, in general, the overall OCS as $\theta_{ij}$ and $\mu$ increase does not simply fade, but the whole FFDP deforms as it is a non-trivial function of the complex electric field and geometry, as expressed by Eq. \eqref{eq:diffraction_integral}. However, this limitation to our model due to this approximation can be partially accounted for through a calibration procedure (see Sec. \ref{sec:model_validation}). 

\subsection{CCR Characterization of Incidence and Velocity Aberration Effects}
As we will now focus on the behavior of a single CCR, we can drop the subscript $i$; moreover, to avoid confusion we will drop also the subscript $j$ as we will never refer to a single LoS, but always to bulk sets of them, indicated by $s$. We will refer to on-design illumination conditions for a CCR, moving with the observer ($\mu$ = 0) when the green ($\lambda = 532$ nm) laser impinges normally ($\theta = 0^\circ$) on the CCR aperture and the first edge is aligned with the positive $x$ direction ($\alpha^{(1)} = 0^\circ$), and both are aligned with the laser polarization direction, which is defined to be horizontal in such case. If not explicitly stated, when a single parameter is changed we imply that the other ones are to be considered fixed under the on-design conditions just described. 
For the present study, several choices of $k^{\mathrm{ab}}(\mu)$ and $k^{\mathrm{inc}}(\theta)$ were considered, including both theoretical and experimentally measured functions. The most faithful results, with respect to the reference values reported in Section \ref{sec:model_validation}, were obtained by using a numerical $k^{\mathrm{inc}}$ curve reconstructed from $63$ different illumination conditions of the CCR, corresponding to 3 azimuths and 21 incidence angles, as reported in Tab. \ref{tab:laser_angles_k_inc_curve}.
\begin{table}[htpb]
    \centering
    \begin{tabular*}{\linewidth}{@{\extracolsep{\fill}} c c c c c @{}}
    \hline
    \textbf{Laser azimuth} & \textbf{1}$^\text{st}$ \textbf{Edge azimuth} & \textbf{Incidence} & \textbf{Index} & \textbf{Weight} \\ 
         \qquad $\alpha$ & $\alpha^{(1)}$ & $\theta$ & $s$ & $w$ \qquad  \\ \hline
         \qquad  $0^\circ$ & $0^\circ$ & $0^\circ : 1^\circ : 21^\circ$ & 1 & 1.0
         \qquad  \\
         \qquad $0^\circ$ & $-60^\circ$ & $0^\circ : 1^\circ : 21^\circ$ & 2 & 2.0
         \qquad \\
         \qquad  $0^\circ$ & $-90^\circ$ & $0^\circ : 1^\circ : 21^\circ$ & 3 & 1.0 
         \qquad   \\ \hline
    \end{tabular*}
    \caption{Orientation of the laser impinging on the test CCR used to retrieve the theoretical $k^{\mathrm{inc}}(\theta)$ curve. The notation $0^\circ : 1^\circ : 21^\circ$ indicates $\theta = \{0^\circ, 1^\circ, 2^\circ, ..., 20^\circ \}$, while $s$ is the subscript associated with the particular illumination condition and $w$ is the weight associated to each curve, as in Eq. \eqref{eq:weighted_avg_k_inc}.}
    \label{tab:laser_angles_k_inc_curve}
\end{table}

\begin{figure}
    \centering
    \includegraphics[width=\linewidth]{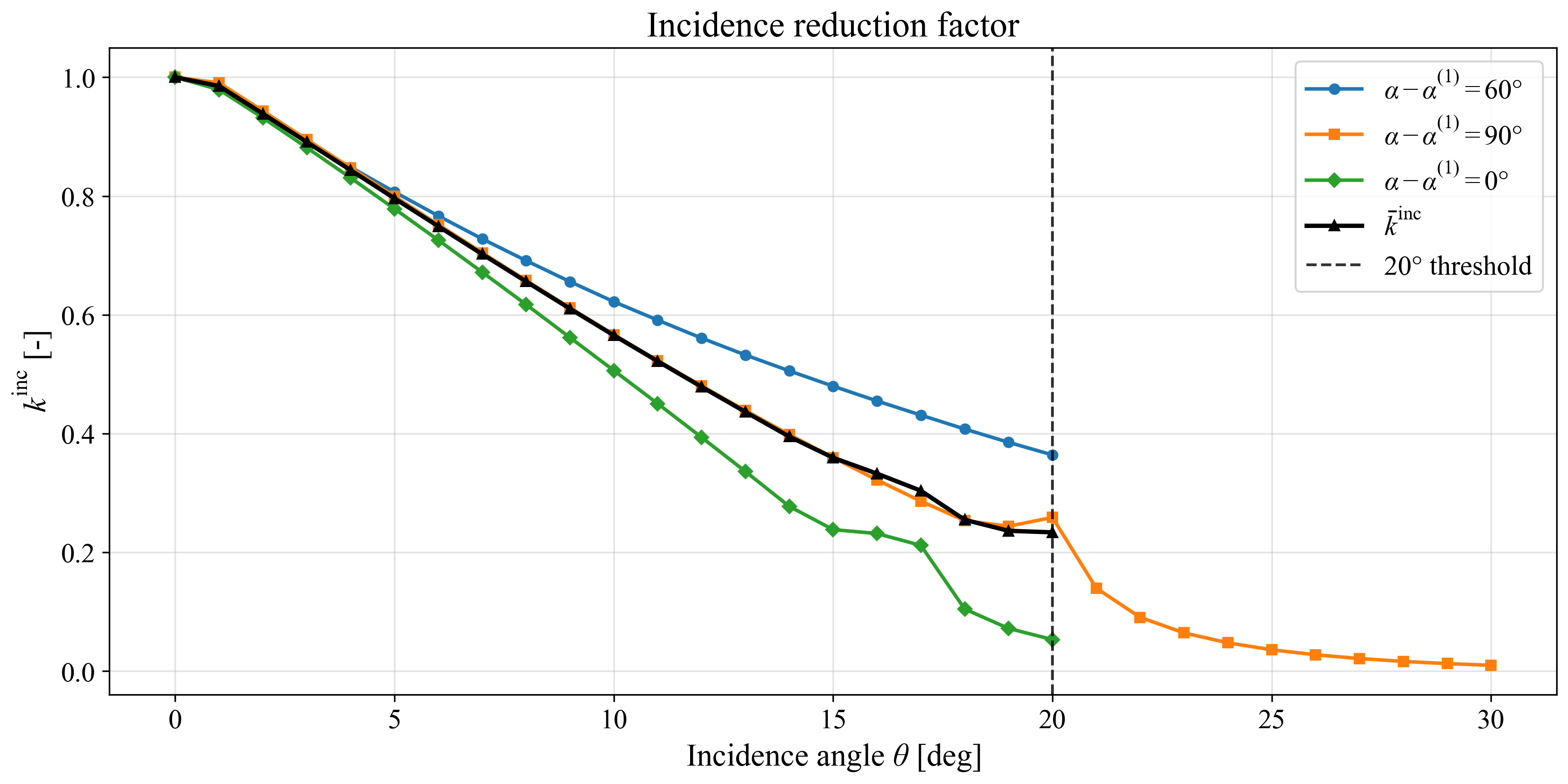}
    \caption{Incidence factor $k^{\mathrm{inc}}(\theta)$ for different azimuthal illumination conditions and weighted average of all the curves.}
    \label{fig:k_inc_curves}
\end{figure}

The resulting curves are shown in Fig.~\ref{fig:k_inc_curves}. The average $\bar k^{\mathrm{inc}}$ curve considered is computed as:
\begin{equation} \label{eq:weighted_avg_k_inc}
\bar k^{\mathrm{inc}}(\theta) = \begin{cases}
\displaystyle
\frac{\sum\limits_{s=1}^{3} (k_s^{\mathrm{inc}}(\theta) \cdot w_s)}{\sum\limits_{j=1}^{3} w_s}, \quad &\theta \leq 20^\circ \\
\displaystyle
0, \quad &\theta > 20^\circ
\end{cases}
\end{equation}
with the weights $w_j$ associated to the $j$-th curve reported in Tab. \ref{tab:laser_angles_k_inc_curve}.

$\bar k^{\mathrm{inc}}(\theta)$ is conservatively truncated for incidence angles $\theta >20^\circ$ for two reasons:
\begin{itemize}
    \item[\ding{226}] To account for unavoidable real-world effects that are not explicitly included in the model, such as surface flatness errors, DAOs tolerances, material impurities, and other manufacturing imperfections inherent to the CCR.
   \item[\ding{226}] Examination of the curve $k_{90^\circ}^{\mathrm{inc}}$ for incidence angles $\theta = 20^\circ : 1^\circ : 31^\circ$ reveals a sharp decrease in the returned signal for $\theta \gtrsim 21^\circ$. In computing the average curve over the interval $\theta \in [0^\circ,20^\circ]$, a weight twice that of the other two curves was assigned to $k_{90^\circ}^{\mathrm{inc}}$. This choice reflects the fact that, during a complete rotation of the CCR ($\alpha = 0^\circ$, $\alpha^{(1)} \in [0^\circ,120^\circ]$, owing to the threefold periodicity discussed previously), this illumination condition occurs twice as frequently as the other two. Owing to the computational cost associated with each simulation, it was not feasible to compute additional curves corresponding to other relative clocking angles $(\alpha-\alpha^{(1)})$. Nevertheless, the results indicate that the curve associated with $k_{0^\circ}^{\mathrm{inc}}$ becomes essentially negligible for $\theta \gtrsim 17^\circ$. Consequently, only the curve corresponding to $k_{60^\circ}^{\mathrm{inc}}$ may still provide a non-negligible contribution. However, it should be noted that this curve is representative only of CCRs oriented at that specific relative clocking angle; therefore, extending its contribution through averaging would almost certainly lead to a wrong overestimation of the returned signal.

\end{itemize}

\begin{figure}[htbp]
    \centering

    \begin{subfigure}[b]{0.75\textwidth}
        \centering
        \includegraphics[width=\linewidth]{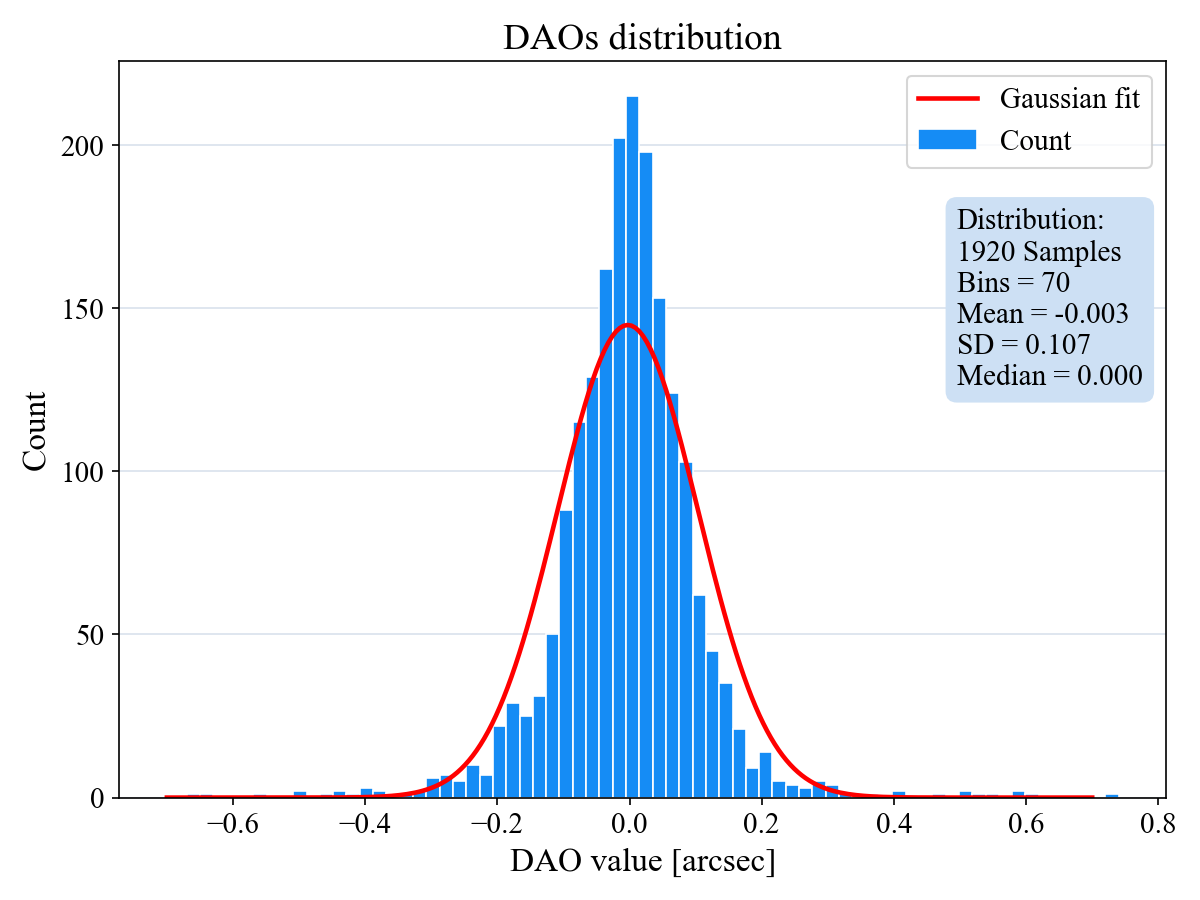}
        \caption{Measured DAOs distribution.}
        \label{fig:dao_distribution_histogram}
    \end{subfigure}
    \hfill
    \begin{subfigure}[b]{0.75\textwidth}
        \centering
        \includegraphics[width=\linewidth]{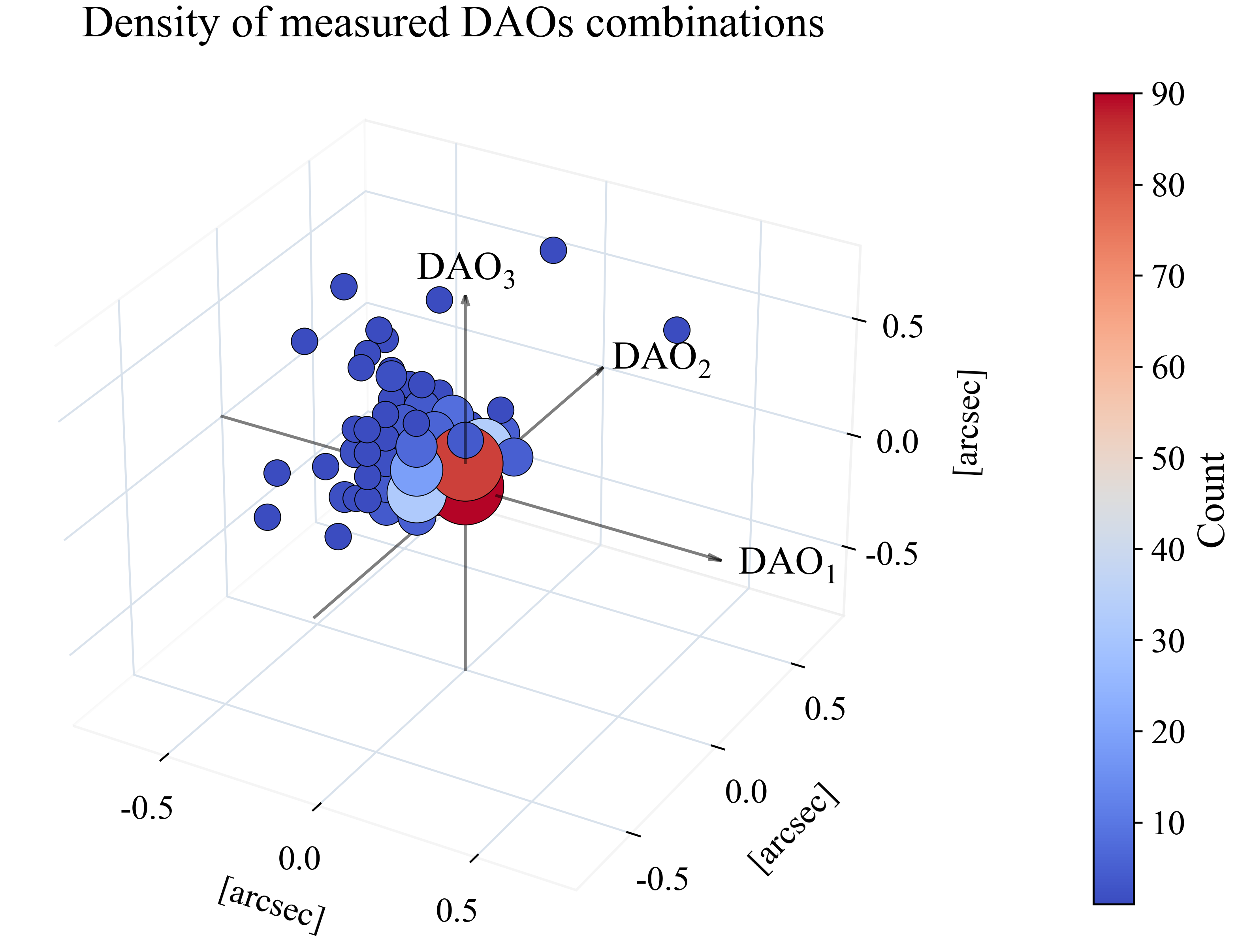}
        \caption{Three-dimensional representation of the most common DAO combinations.}
        \label{fig:dao_distribution_3d}
    \end{subfigure}

    \caption{DAO characterization of the LARES-2 CCR batch: (a) measured DAO distribution for 640 CCRs, and (b) three-dimensional visualization of the most frequent CCR types, categorized according to their DAO combinations.}
    \label{fig:dao_distribution}
\end{figure}

The $k^{\mathrm{ab}}(\mu)$ curve used in this work was also obtained through a weighted-average model. In this case, however, the curve is not entirely theoretical as it is based on experimental measurements as well. Manufactured CCRs exhibit tolerances that affect the FFDP, and these tolerances produce DAOs among the three back faces of the CCR. DAOs can be included in numerical simulations and measured in the laboratory by interferometry. During the construction of LARES-2, a measurement campaign was performed on the full batch of 640 $1\,\mathrm{inch}$ CCRs, nominally with sets of three null DAOs within $\pm 0.86\,\mathrm{arcsec}$, as reported in Fig.~\ref{fig:dao_distribution}.

The 11 most frequent CCR types, categorized according to their DAOs, are reported in Tab.~\ref{tab:dao_combo_counts}. The $k^{\mathrm{ab}}(\mu)$ curve used for the reduced-order modeling of the satellite OCS was computed by running simulations for each of these CCR types and then taking the weighted average of the resulting curves. The weight assigned to each curve was proportional to the occurrence count of the corresponding DAO combination in the measured population. For the sake of completeness, we highlight that we performed the optical simulations, retrieving the FFDPs associated with each set of DAOs, only for on-design illumination conditions.

Following a similar definition of the one used for $\bar k^{\mathrm{inc}}(\theta)$, in Eq. \eqref{eq:weighted_avg_k_ab} we can define:

\begin{equation} \label{eq:weighted_avg_k_ab}
\bar k^{\mathrm{ab}}(\mu) = \displaystyle
\frac{\sum\limits_{d=1}^{11} (k_d^{\mathrm{ab}}(\mu) \cdot w_d)}{\sum\limits_{d=1}^{11} w_d},
\end{equation}

with no truncation needed this time.

\begin{figure}[htbp]
    \centering

    \begin{subfigure}[b]{\textwidth}
        \centering
        \includegraphics[width=\linewidth]{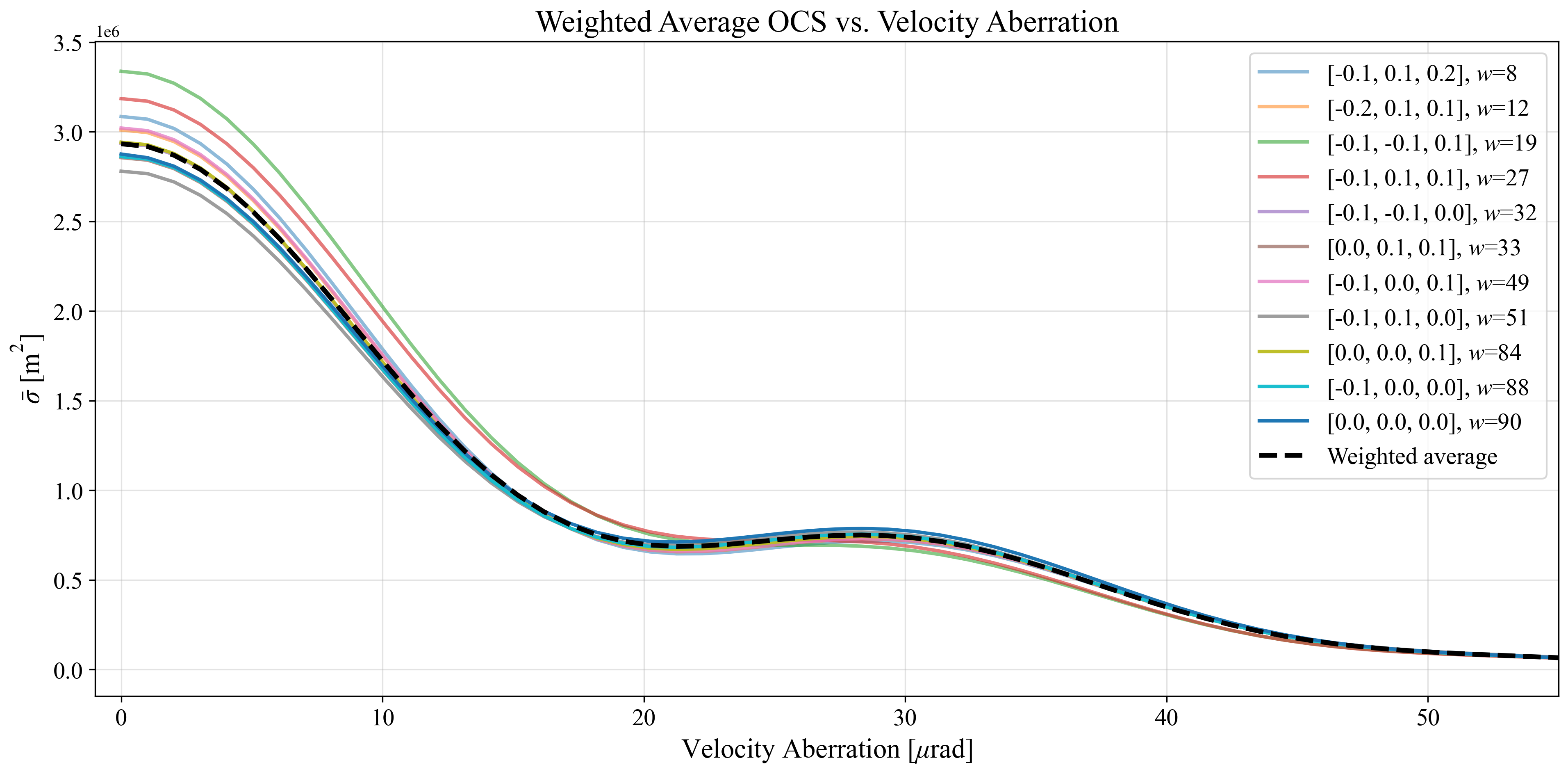}
        \caption{Individual simulated curves.}
        \label{fig:k_ab_individual_daos}
    \end{subfigure}
    \hfill
    \begin{subfigure}[b]{\textwidth}
        \centering
        \includegraphics[width=\linewidth]{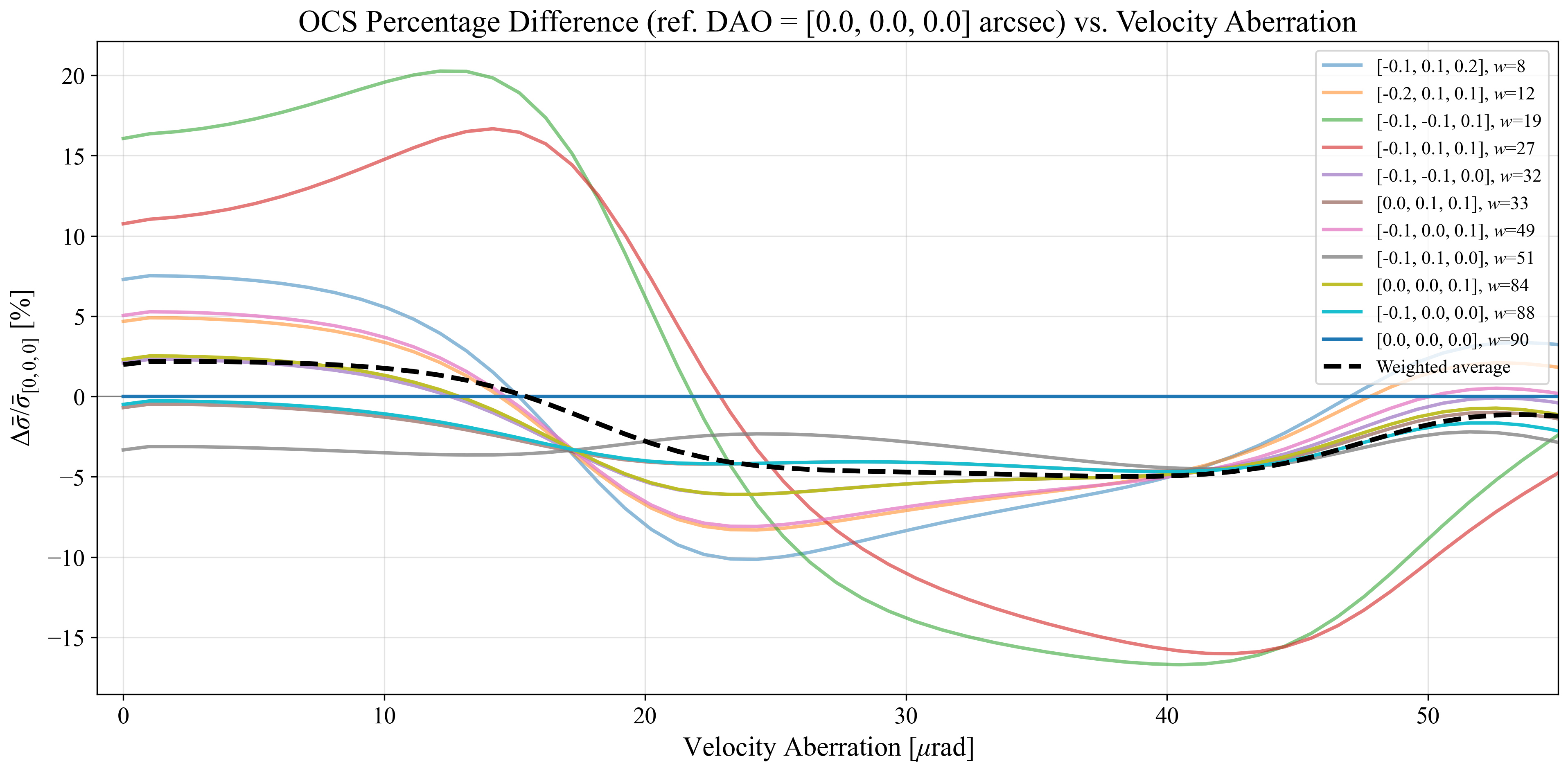}
        \caption{Percentage differences curves.}
        \label{fig:k_ab_weighted_average}
    \end{subfigure}

    \caption{Simulated $\sigma(\mu)$ curves for the considered DAO combinations: (a) individual curves and (b) corresponding percentage differences curves with respect to the null DAOs CCR.}
    \label{fig:k_ab_weighted}
\end{figure}

The resulting curves are shown in Fig.~\ref{fig:k_ab_individual_daos}. By taking the curve with three equal null DAOs as a reference, and computing the percentage difference between it and the other curves (as displayed in Fig.\ref{fig:k_ab_weighted_average}) we can see how the error between the $k^{\mathrm{ab}}_{[0,0,0]}(\mu)$ and $\bar k^{\mathrm{ab}}(\mu)$ is always within $\pm 5\%$. For this reason, we will not induce a large error by approximating the curve $\bar k^{\mathrm{ab}}(\mu)$ with the one relative to the null DAOs case, as will be done for the CCR2 population. This actually holds considering that GUEST aims at having non-COTS (Commercial Off The Shelf), tailored CCRs, which benefit from stricter manufacturing tolerances.

\begin{table}[htbp]
\centering
\begin{tabular*}{0.75\linewidth}{@{\extracolsep{\fill}}ccccc}
\hline
\multicolumn{3}{c}{\textbf{DAOs} [arcsec]} & \textbf{Count}  & \textbf{Index}\\
% \cline{1-3}
DAO$_1$ & DAO$_2$ & DAO$_3$ & $w$ & $d$\\
\hline
$\left(\right.$ 0.00, &  0.00, & 0.00$\left.\right)$ & 90 & 1\\
$\left(\right.$-0.10, &  0.00, & 0.00$\left.\right)$ & 88 & 2\\
$\left(\right.$ 0.00, &  0.00, & 0.10$\left.\right)$ & 84 & 3\\
$\left(\right.$-0.10, &  0.10, & 0.00$\left.\right)$ & 51 & 4\\
$\left(\right.$-0.10, &  0.00, & 0.10$\left.\right)$ & 49 & 5\\
$\left(\right.$ 0.00, &  0.10, & 0.10$\left.\right)$ & 33 & 6\\
$\left(\right.$-0.10, & -0.10, & 0.00$\left.\right)$ & 32 & 7\\
$\left(\right.$-0.10, &  0.10, & 0.10$\left.\right)$ & 27 & 8\\
$\left(\right.$-0.10, & -0.10, & 0.10$\left.\right)$ & 19 & 9\\
$\left(\right.$-0.20, &  0.10, & 0.10$\left.\right)$ & 12 & 10\\
$\left(\right.$-0.10, &  0.10, & 0.20$\left.\right)$ & 8 & 11\\
\hline
\end{tabular*}
\caption{Counts of the most frequent DAOs combinations in the LARES-2 measured batch of 640 CCRs with a binning of 0.1 arcsec.}
\label{tab:dao_combo_counts}
\end{table}

\section{Validation and calibration of the model}
\label{sec:model_validation}

\begin{figure}
    \centering
    \includegraphics[width=0.75\linewidth, trim = 0 3.5cm 0 0, clip]{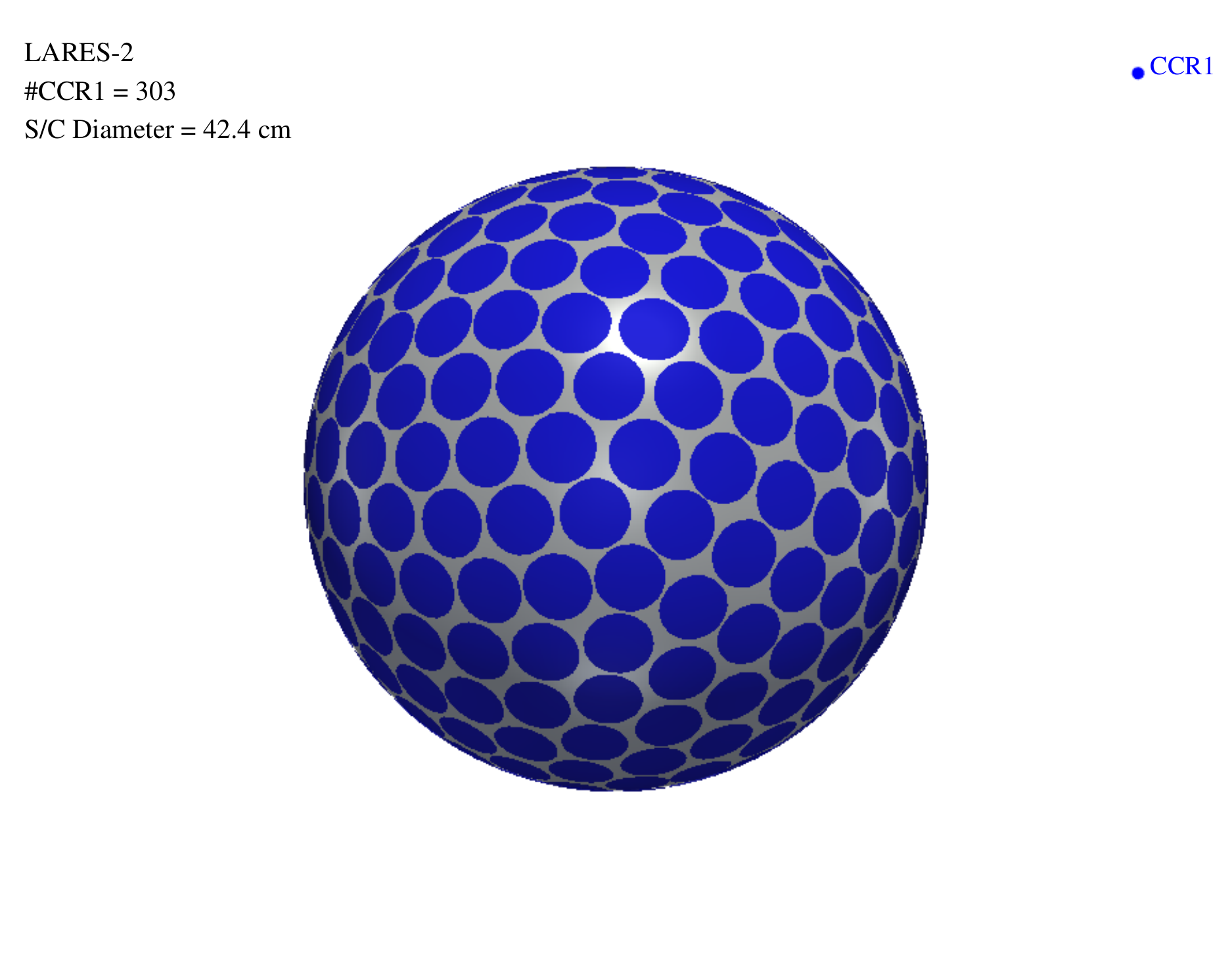}
    \caption{Reconstructed geometrical model of the LARES-2 satellite.}
    \label{fig:lares_2_reconstructed}
\end{figure}

In order to assess the validity of the proposed methodology for estimating the total satellite OCS, the LARES-2 model was reconstructed with 303 CCR1s arranged according to the Thomson solution~\cite{ciufolini_lares2,arnold_lares_transfer_2015}, as shown in Fig.~\ref{fig:lares_2_reconstructed}. The total OCS is known from previous analyses performed with accurate software. However, applying such tools at this early stage of the design process would be impractical, especially for the sensitivity analyses considered here. For the real as-built satellite, the average OCS was estimated to be $\sigma = 2.7 \times 10^6\,\mathrm{m}^2$~\cite{ciufolini_lares2,pirrotta2025}. Given the low eccentricity of the LARES-2 orbit ($e = 0.00027$)~\cite{pirrotta2025} a constant velocity aberration of $\mu_{\mathrm{LARES-2}} = 38.0\,\mu\mathrm{rad}$ was assumed. For the spherical-cap modeling of the CCR1 population, the CCR housing thickness and the clearance distance between adjacent mounted housings, were derived from technical documentation of the actual mounting setup. In the final as-built satellite, the arrangement was locally modified to accommodate the four holes required for the Hold Down \& Release Mechanism (HDRM). This local modification was deemed unimportant and was neglected in the LARES-2 reconstruction; it will also not be included in the GUEST modeling at this phase.

The model described in Section~\ref{sec:reduced_order_model} was applied using the set of LoS defined by the angles reported in Tab.~\ref{fig:los_distribution_table}. The total number of LoS is set to 182 as tests performed with a larger number of LoS, did not yield appreciable differences.

\newlength{\subfigheight}
\setlength{\subfigheight}{7.0cm} % choose one common height

\begin{figure}[htbp]
    \centering
    \begin{subfigure}[c]{0.44\textwidth}
        \centering
        \begin{minipage}[t][\subfigheight][b]{\linewidth}
            \centering
            \vspace{1.25cm}
            \begin{tabular*}{\linewidth}{@{\extracolsep{\fill}} l l c @{}}
            \hline
            \multicolumn{1}{c}{S/C Latitude} &
            \multicolumn{1}{c}{S/C Azimuth} &
            \multicolumn{1}{c}{LoS per ring} \\
            \multicolumn{1}{c}{$\psi_{S/C}$} &
            \multicolumn{1}{c}{$\omega_{S/C}$} &
            \multicolumn{1}{c}{} \\
            \hline
             $-90.0^\circ$ & $0^\circ$ & 1  \qquad \\
             $-60.0^\circ$ & $0^\circ:18^\circ:360^\circ$  & 20 \qquad \\
             $-30.0^\circ$ & $0^\circ:9^\circ:360^\circ$   & 40 \qquad \\
             $+0.0^\circ$  & $0^\circ:6^\circ:360^\circ$   & 60 \qquad \\
             $+30.0^\circ$ & $0^\circ:9^\circ:360^\circ$   & 40 \qquad \\
             $+60.0^\circ$ & $0^\circ:18^\circ:360^\circ$  & 20 \qquad \\
             $+90.0^\circ$ & $0^\circ$ & 1  \qquad \\
            \hline
            \end{tabular*}

            \vfill

            \caption{}
            \label{fig:los_distribution_table}
        \end{minipage}
    \end{subfigure}
    \hfill
    \begin{subfigure}[c]{0.55\textwidth}
        \centering
        \begin{minipage}[t][\subfigheight][b]{\linewidth}
            \centering

            \tdplotsetmaincoords{70}{210}

            \begin{tikzpicture}[
                tdplot_main_coords,
                scale=2.5,
                los/.style={-{Latex[length=2mm,width=1.4mm]}, very thick, ForestGreen!90}
            ]
                \input{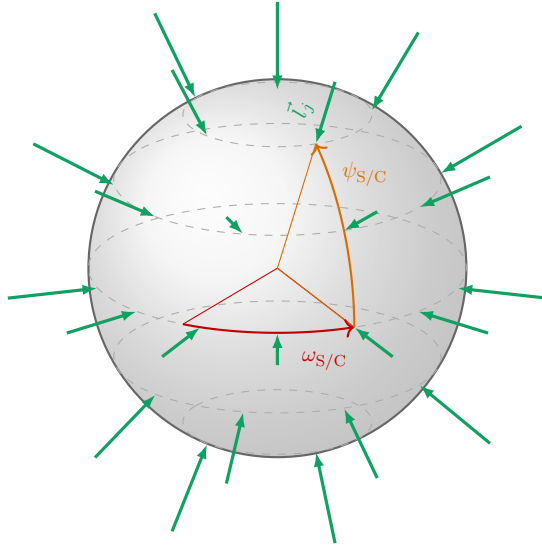}
            \end{tikzpicture}

            \vfill

            \caption{}
            \label{fig:los_unit_sphere}
        \end{minipage}
    \end{subfigure}

    \caption{Angular definition of the LoS used to compute the LARES-2 OCS and corresponding distribution of LoS vectors impinging on the unit sphere.}
    \label{fig:los_distribution_combined}
\end{figure}

Using the curves $\bar k^{\mathrm{inc}}(\theta_{ij})$ and $\bar k^{\mathrm{ab}}(\mu)$ described in the previous Section, applying Eq. \eqref{eq:total_sc_ocs} to the LoS defined in Tab. \ref{fig:los_distribution_table}, the resulting OCS estimate for the LARES-2 satellite yields
\begin{equation} \label{eq:lares2_tot_ocs}
    \bar \sigma_{\mathrm{LARES\text{-}2}}
    = 
\frac{1}{N_{\mathrm{LoS}}}
\sum_{j=1}^{N_{\mathrm{LoS}}}
\bar \sigma_{\mathrm{LARES-2}, j}(\mu_{\mathrm{LARES-2}}) = 
    (1.89 \pm 0.09)\times 10^6\,\mathrm{m}^2.
\end{equation}
This value is deemed to be in reasonable agreement with the expected LARES-2 OCS of $2.7 \times 10^6\,\mathrm{m}^2$, despite being approximately $30\%$ lower. The underestimation can be attributed to the simplified treatment of the effects introduced by clocking the CCR configuration adopted in the present model. In fact, optimized CCR clocking has been shown to provide an overall increase in OCS by amounts consistent with this discrepancy~\cite{Degnan2023}. This suggests that this missing contribution may be effectively recovered in GUEST and LARES models by appropriately clocking the CCRs on the satellite surface. Conversely, for a randomly clocked CCR configuration, an average value close to $(1.89 \pm 0.09)\times10^6\,\mathrm{m}^2$ would be expected. The calibration factor for the proposed methodology therefore is:
\begin{equation}
    C_\mathrm{LARES-2}
    =
    \frac{2.7}{1.89}
    \simeq
    1.42857.
    \label{eq:calibration_factor}
\end{equation}

\section{GUEST baseline configuration}
\label{sec:guest_baseline_configuration}

The baseline GUEST configuration is defined as the reference opto-geometrical layout against which the sensitivity analyses are performed. It is based on the heritage of dense passive laser-ranged satellites, such as LAGEOS and LARES~\cite{ILRS_LARES, ILRS_LAGEOS}, but it is adapted to the much more demanding GUEST orbital regime (as defined in Tab. \ref{fig:orbital_parameters_table}), which in turn translates to a large variation of VA between perigee and apogee observing conditions (Fig. \ref{fig:VA_vs_altitude}). 

\setlength{\subfigheight}{7cm}

\begin{figure}[htbp]
    \centering

    \begin{subfigure}[c]{0.48\textwidth}
        \centering
        \begin{minipage}[t][\subfigheight][b]{\linewidth}
            \centering
            \small

            \textbf{Common orbital parameters:}

            \vspace{0.15cm}

            \begin{tabular*}{\linewidth}{@{\extracolsep{\fill}} l l @{}}
            \hline
            \textbf{Parameter} & \textbf{Value} \\
            \hline
            Perigee altitude & 6 926 km \\
            Apogee altitude & 86 700 km \\
            Eccentricity & 0.75 \\
            Period & 33.9 h \\
            Argument of perigee & 270$^\circ$ \\
            Semi-major axis & 53 000 km \\
            \hline
            \end{tabular*}

            \vspace{0.3cm}

            \textbf{Spacecraft parameters:}

            \vspace{0.15cm}

            \begin{tabular*}{\linewidth}{@{\extracolsep{\fill}} l c c c @{}}
            \hline
            \textbf{Satellite} & \textbf{Inclination} & \textbf{RAAN} & \textbf{Reentry time} \\
            \hline
            GUEST-1 & 75$^\circ$ & 0$^\circ$ & 29.5 yr \\
            GUEST-2 & 50$^\circ$ & 220$^\circ$ & 32.4 yr \\
            \hline
            \end{tabular*}
            \vspace{0.75cm}
            \caption{}
            \label{fig:orbital_parameters_table}
        \end{minipage}
    \end{subfigure}
\hfill
    \begin{subfigure}[c]{0.48\textwidth}
        \centering
        \begin{minipage}[t][\subfigheight][b]{\linewidth}
            \centering
            \includegraphics[width=\linewidth]{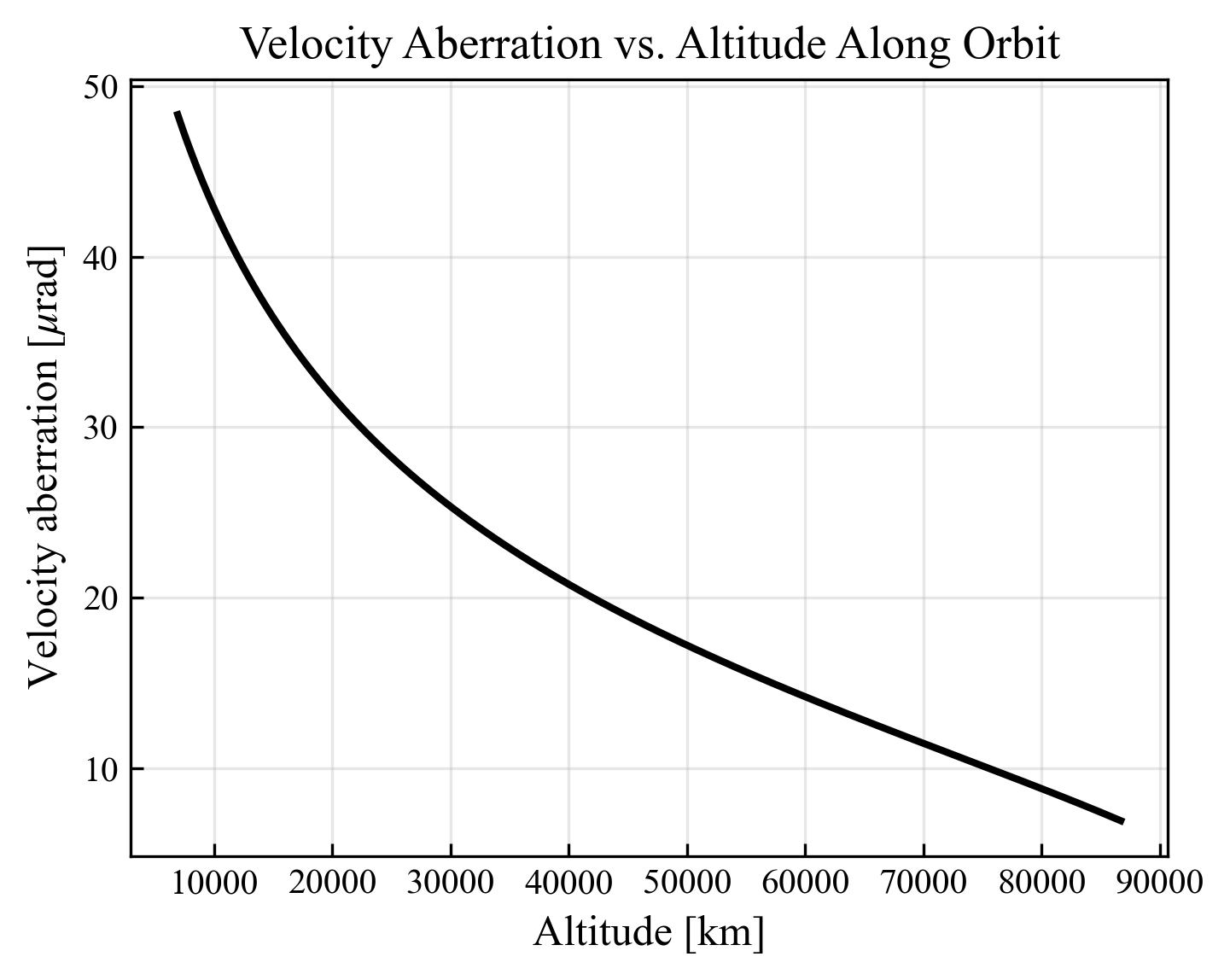}
            \vfill
            \caption{}
            \label{fig:VA_vs_altitude}
        \end{minipage}
    \end{subfigure}

    \caption{GUEST orbital parameters for the two proposed orbits (a) and Velocity Aberration vs. Altitude  along the orbit, computed through the formula reported in Eq. \eqref{eq:velocity_aberration} (b).}
    \label{fig:VA_and_orbital_table}
\end{figure}

Each variation considered below changes only one key parameter with respect to this reference case, as summarized in Tab.~\ref{tab:guest_configurations}. Of course, since illumination at a different wavelength is possible for the same configuration, the ``Infrared'' case should be interpreted as a wavelength sensitivity study rather than a separate spacecraft design.

\begin{figure}[htbp]
    \centering

    \begin{subfigure}[b]{0.48\textwidth}
        \centering
        \includegraphics[width=\linewidth, trim=0 3.5cm 0 0, clip]{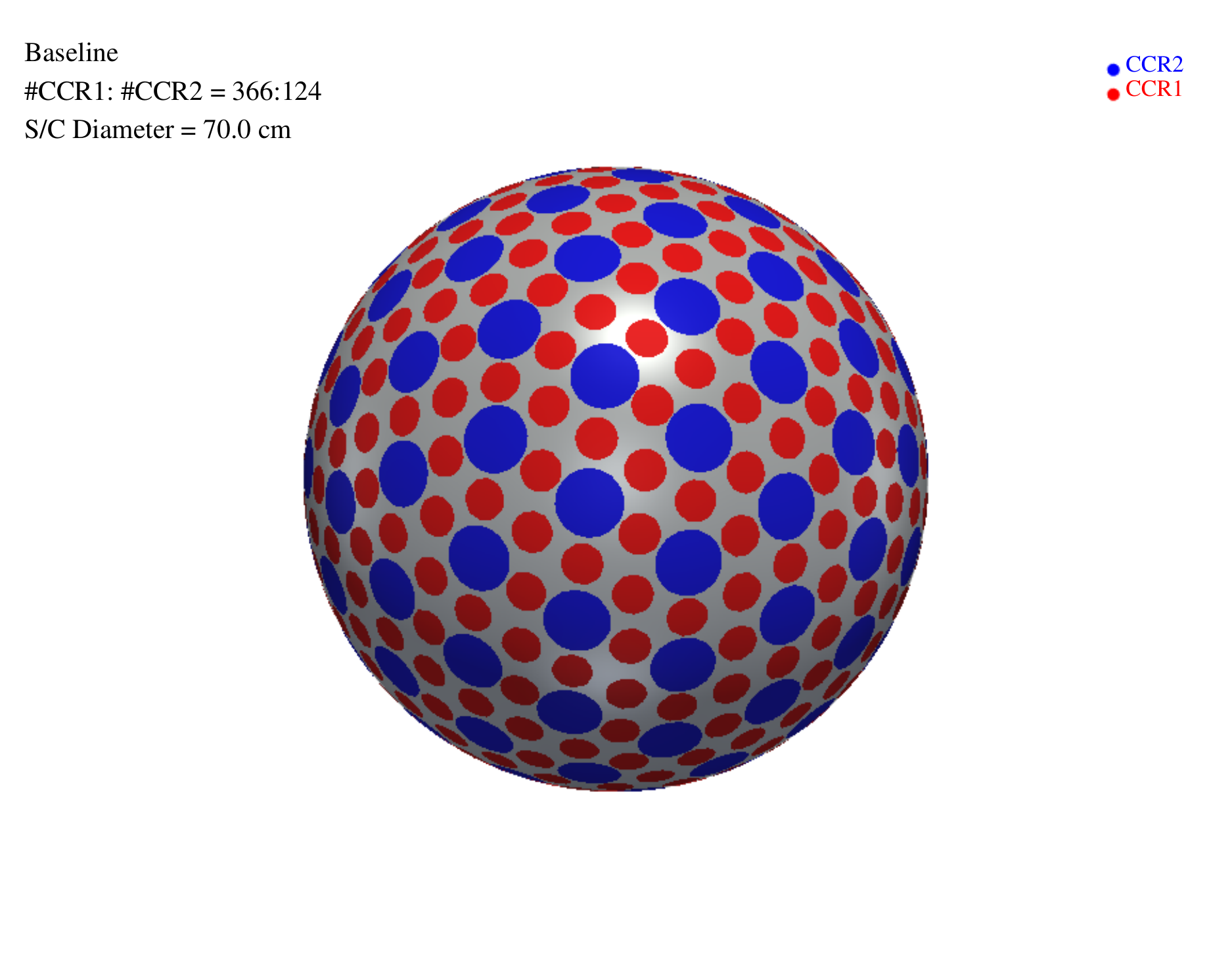}
        \caption{GUEST baseline configuration.}
        \label{fig:guest_baseline_350mm_31}
    \end{subfigure}
    \hfill
    \begin{subfigure}[b]{0.48\textwidth}
        \centering
        \includegraphics[width=\linewidth, trim=0 3.5cm 0 0, clip]{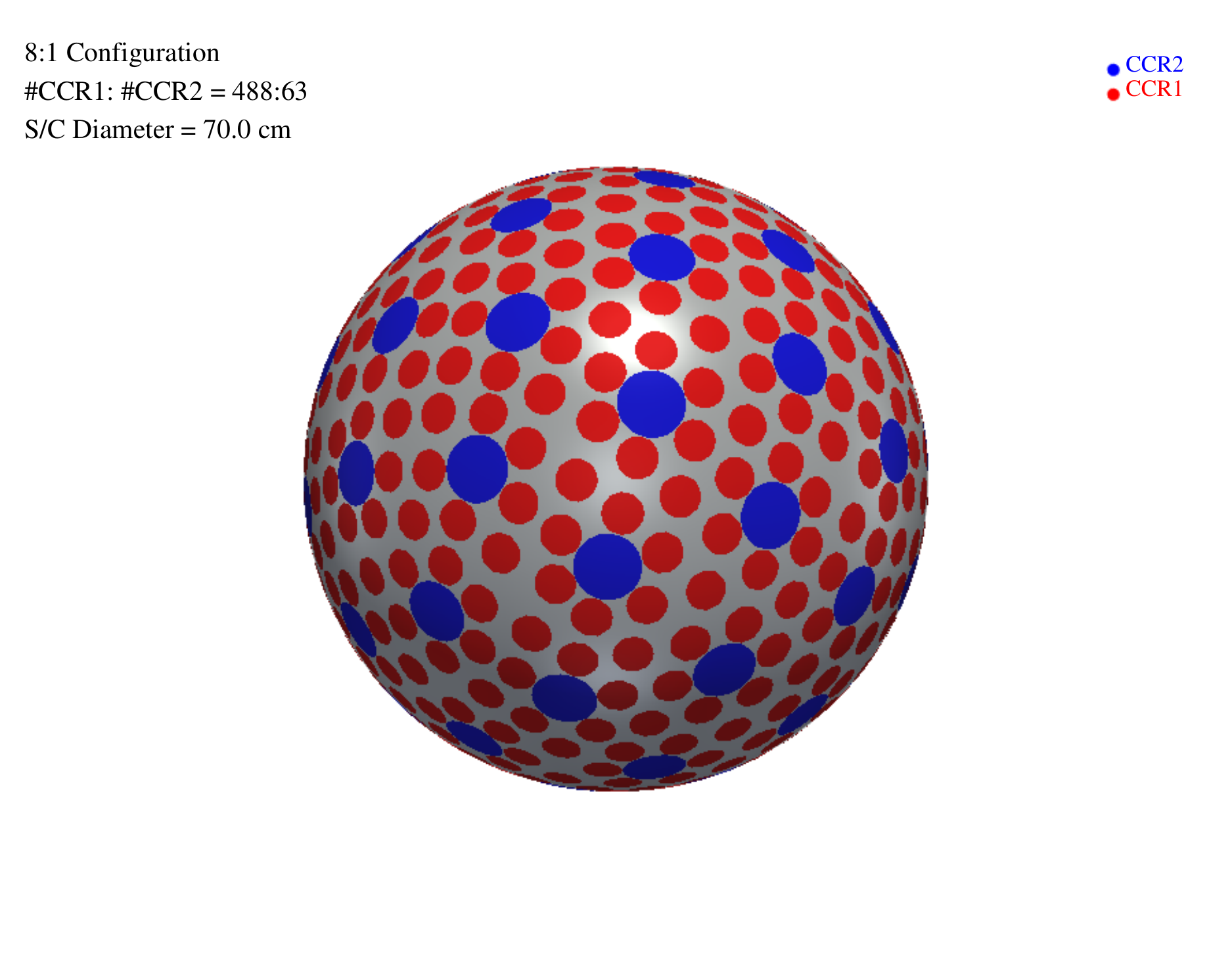}
        \caption{GUEST $8{:}1$ configuration.}
        \label{fig:guest_8to1_350mm}
    \end{subfigure}

    \vspace{0.5cm}

    \begin{subfigure}[b]{0.48\textwidth}
        \centering
        \includegraphics[width=\linewidth, trim=0 3.5cm 0 0, clip]{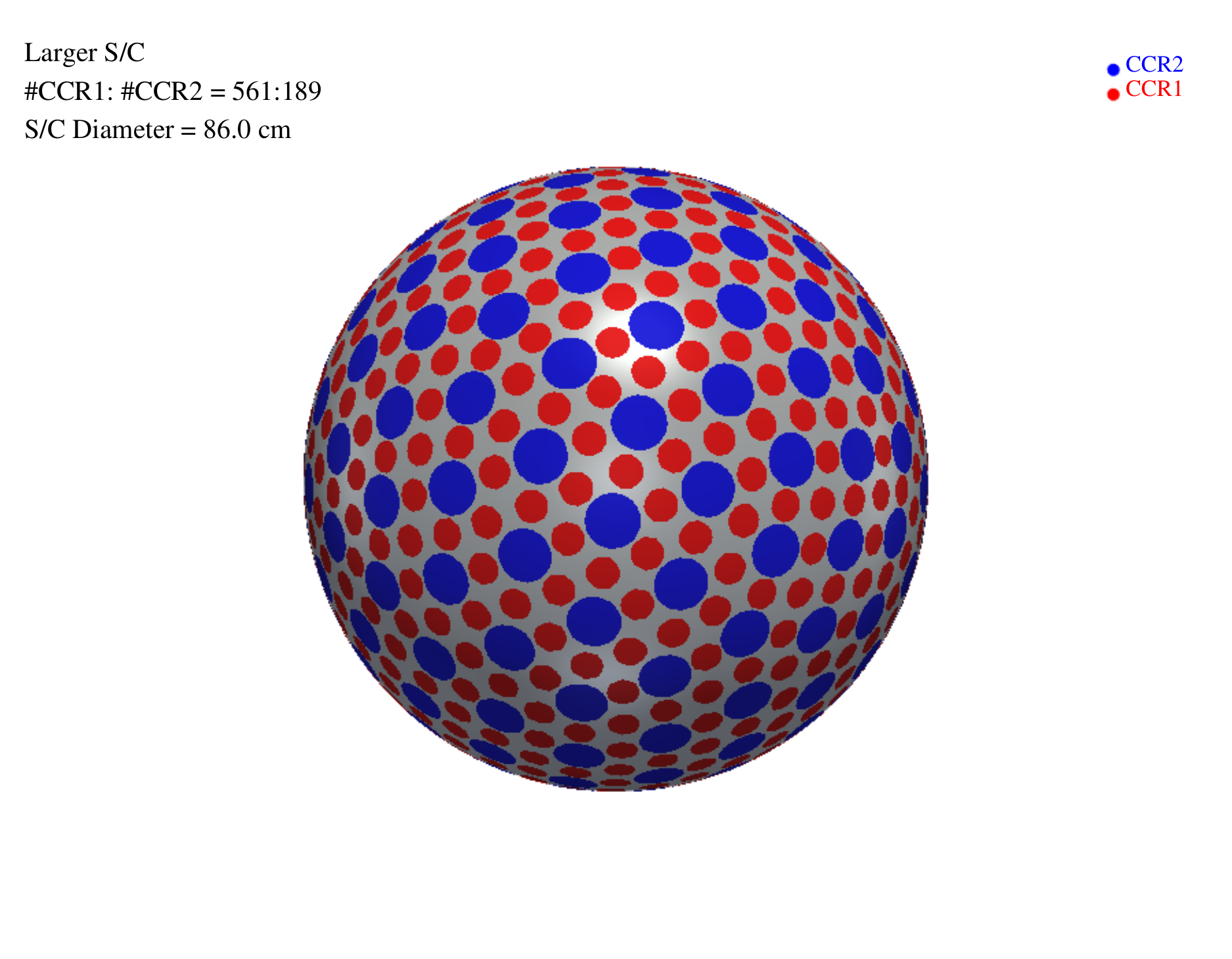}
        \caption{GUEST $R_{\mathrm{S/C}} = 43\,\mathrm{cm}$.}
        \label{fig:guest_larger_sc}
    \end{subfigure}
    \hfill
    \begin{subfigure}[b]{0.48\textwidth}
        \centering
        \includegraphics[width=\linewidth, trim=0 3.5cm 0 0, clip]{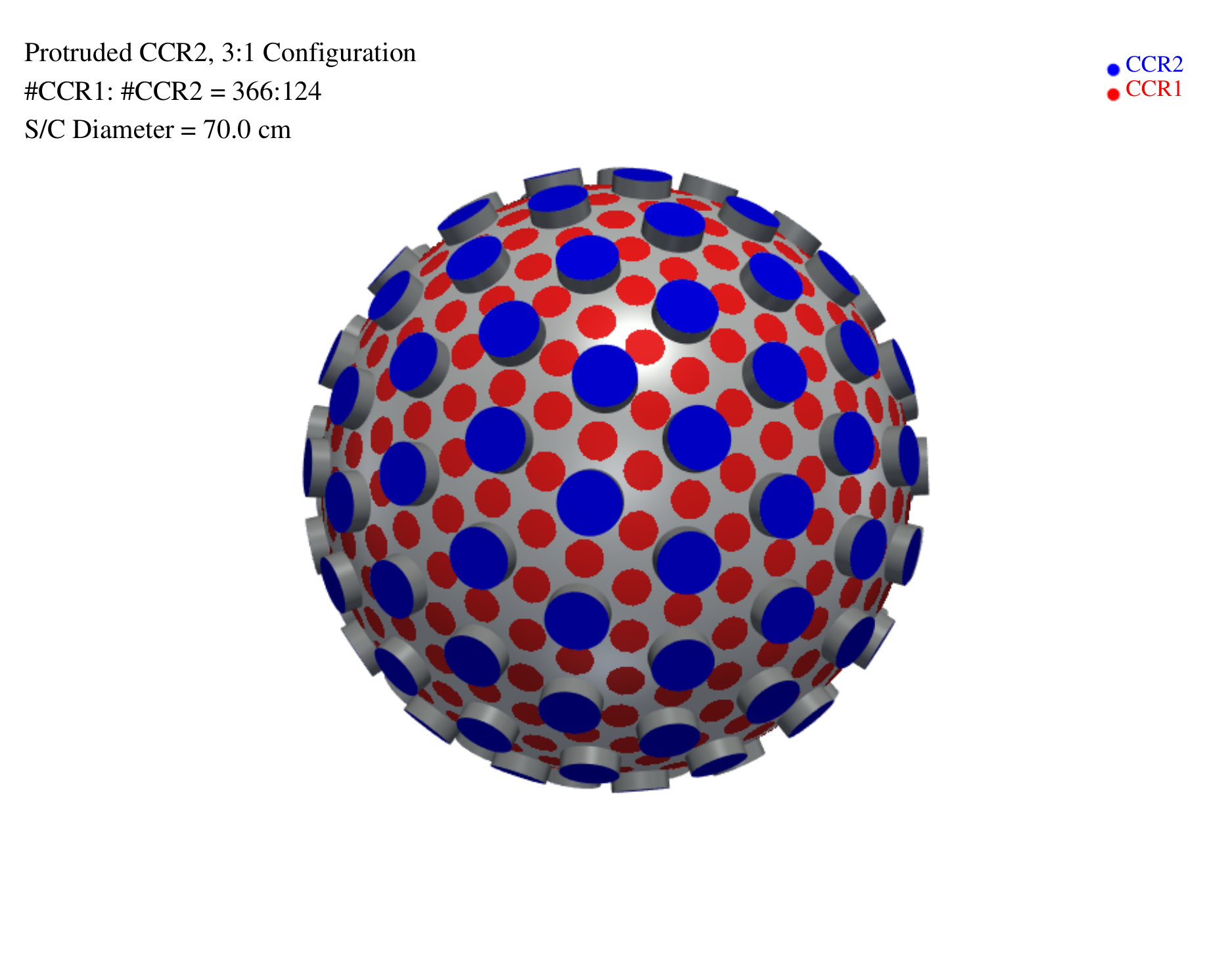}
        \caption{GUEST baseline mechanical configuration.}
        \label{fig:protruded_ccr2_31_config}
    \end{subfigure}

    \caption{Comparison of the GUEST spacecraft configurations considered in the analysis.}
    \label{fig:guest_configurations_grid}
\end{figure}

A rendering of the baseline configuration is shown in Fig.~\ref{fig:guest_baseline_350mm_31}.

\begin{table}[htpb]
\centering
\renewcommand{\arraystretch}{1.25}
\begin{tabular*}{\linewidth}{@{\extracolsep{\fill}}cccc}
\hline
\textbf{Configuration} &
\textbf{Parameters} &
\textbf{Perigee OCS} &
\textbf{Apogee OCS} \\
&&$\bar \sigma_{\mathrm{GUEST}}(\mu_{p})$ & $\bar \sigma_{\mathrm{GUEST}}(\mu_{a})$ \\
\hline
Baseline &
\begin{tabular}[c]{@{}c@{}}
$3{:}1$ CCR1/CCR2 ratio;\\
$R_{\mathrm{S/C}}=35\,\mathrm{cm}$;\\
$\lambda=532\,\mathrm{nm}$;\\
DAOs $=(0,0,0)\,\mathrm{arcsec}$
\end{tabular}
&
1.02 &
31.82 \\ 
\multicolumn{4}{c}{\rule[0.5ex]{0pt}{0pt}
- - - - - - - - - - - - - - - - - - - - - - - - - - - - - - - - - - - - - - - - - - - - - - - - - - - - - - - - - - - } \\

Infrared &
$\lambda=1064\,\mathrm{nm}$ &
1.66 (+62.7\%) &
13.28 (-58.3\%) \\

$8{:}1$ configuration &
$8{:}1$ CCR1/CCR2 ratio &
1.35 (+32.4\%) &
22.89 (-28.1\%) \\

Larger satellite &
$R_{\mathrm{S/C}}=43\,\mathrm{cm}$ &
1.55 (+52.0\%) &
48.60 (+52.8\%) \\

Non-null DAOs &
CCR1 DAOs $=(1.9,1.9,1.9)\,\mathrm{arcsec}$ &
1.44 (+41.2\%) &
24.59 (-22.7\%) \\
\hline
\end{tabular*}
\caption{Summary of the opto-geometrical configurations analysed for the GUEST dual-CCR satellite. OCS values are reported in units of $10^6\,\mathrm{m}^2$.}
\label{tab:guest_configurations}
\end{table}

The following subsections summarize the advantages and disadvantages of each configuration variant.

\subsection{Infrared}
\label{sec:infrared_configuration}

Changing the laser wavelength modifies the angular scale of the FFDP (Fig. \ref{fig:ffdp_ocs_vs_va_panels}) with a dependency in the ratio $\lambda / D$ ~\cite{Degnan2023}. For a fixed CCR aperture, the diffraction pattern broadens when $\lambda$ is increased~\cite{Delva2023}. As a consequence, the OCS sampled at high velocity aberration can increase, because a larger fraction of the diffracted intensity is displaced toward larger values of $\mu$. This explains the increase of $\bar \sigma_{\mathrm{GUEST}}(\mu_p)$ in the infrared case.

The same effect is not favourable at the apogee. At small velocity aberration, the relevant part of the response is close to the central region of the FFDP, where the shorter wavelength produces a more concentrated pattern. Therefore, the infrared case improves the perigee OCS but significantly reduces the apogee OCS. This makes $\lambda=1064\,\mathrm{nm}$ useful as a perigee-oriented sensitivity case, rather than as a uniformly better solution for the full orbit.

\subsection{$8{:}1$ configuration}
\label{sec:eight_to_one_configuration}

As discussed previously, different CCR1/CCR2 ratios can be obtained for which the satellite's design satisfies the isotropy and homogeneity requirements, provided that the corresponding distribution can be fitted within a triangular tile as described in Sec.~\ref{sec:cap_arrangement}. The $8{:}1$ configuration increases the relative number of CCR1s with respect to the baseline. The purpose is to increase the return at perigee, where the velocity aberration is larger and where the CCR1 population is more efficient.

The result is a higher $\bar \sigma_{\mathrm{GUEST}}(\mu_p)$, with an increase of $32.4\%$ with respect to the baseline. This improvement is obtained at the cost of reducing the CCR2 population, and therefore the apogee OCS decreases by $28.1\%$. The $8{:}1$ configuration is thus a perigee-biased solution. It may be useful if the construction of normal points close to perigee becomes the dominant optical requirement, but it is less balanced than the $3{:}1$ baseline when considering the entire orbit.

A rendering of the sphere produced by this configuration is shown in Fig.~\ref{fig:guest_8to1_350mm}.

\subsection{Larger satellite}
\label{sec:larger_satellite_configuration}

The larger-satellite configuration increases the spacecraft radius from $35\,\mathrm{cm}$ to $43\,\mathrm{cm}$. This variation increases the available spherical surface and therefore allows a larger effective optical payload, while preserving the same optical logic of the baseline configuration. Unlike the infrared, $8{:}1$, and non-null-DAO cases, this modification does not selectively optimize one velocity-aberration regime. Instead, it increases the OCS both at perigee and at apogee.

This is reflected in Tab.~\ref{tab:guest_configurations}: $\bar \sigma_{\mathrm{GUEST}}(\mu_p)$ increases by $52.0\%$, while $\bar \sigma_{\mathrm{GUEST}}(\mu_a)$ increases by $52.8\%$. The gain is therefore approximately uniform for the two reference orbital regions. The drawback is not optical, but mechanical and programmatic. A larger satellite implies a larger mass, larger machining requirements, larger launch and separation-system constraints, and a higher cost. Nevertheless, since GUEST is a passive and mechanically simple satellite, this solution remains attractive if it can be accommodated within the launcher and mission constraints.

However, from a scientific point of view a larger satellite also implies a greater surface-to-mass ratio, enabling the minimization of non-gravitational perturbations~\cite{ciufolini_lares2,arnold_thermal_optical_2020}. A rendering of this configuration is shown in Fig.~\ref{fig:guest_larger_sc}.

\subsection{Non-null DAOs}
\label{sec:non_null_daos}

DAOs, often referred to in the literature as spoiling angles, can be introduced to displace and reshape the FFDP in order to maximize the OCS around a desired velocity aberration~\cite{Degnan2023,Otsubo2011,Williams_2023}.

To maintain the highest possible degree of homogeneity in the present reduced-order model, the main analysis was restricted to sets of three equal DAOs. Allowing three different DAO values would introduce additional azimuthal structure in the FFDP~\cite{Otsubo2011} and increase the dependence of the response on the relative clocking between the CCR and the VA direction. Since these effects are not fully captured by the current model, the corresponding results should be regarded as indicative rather than predictive. Nevertheless, they remain useful for a qualitative assessment of the potential benefits of DAOs inclusion in the CCRs populations.

The value of the equal-DAO set was determined iteratively for the CCR1 population, because the objective is to maximize the OCS at perigee for the construction of normal points. FFDP simulations were performed for CCRs with equal DAOs in the range $0:0.1:2.3\,\mathrm{arcsec}$. The configuration with DAOs equal to 1.9$\,\mathrm{arcsec}$ was found to be particularly favourable for the perigee velocity aberration.

\begin{figure}[htbp]
\centering
        \includegraphics[width=\textwidth]{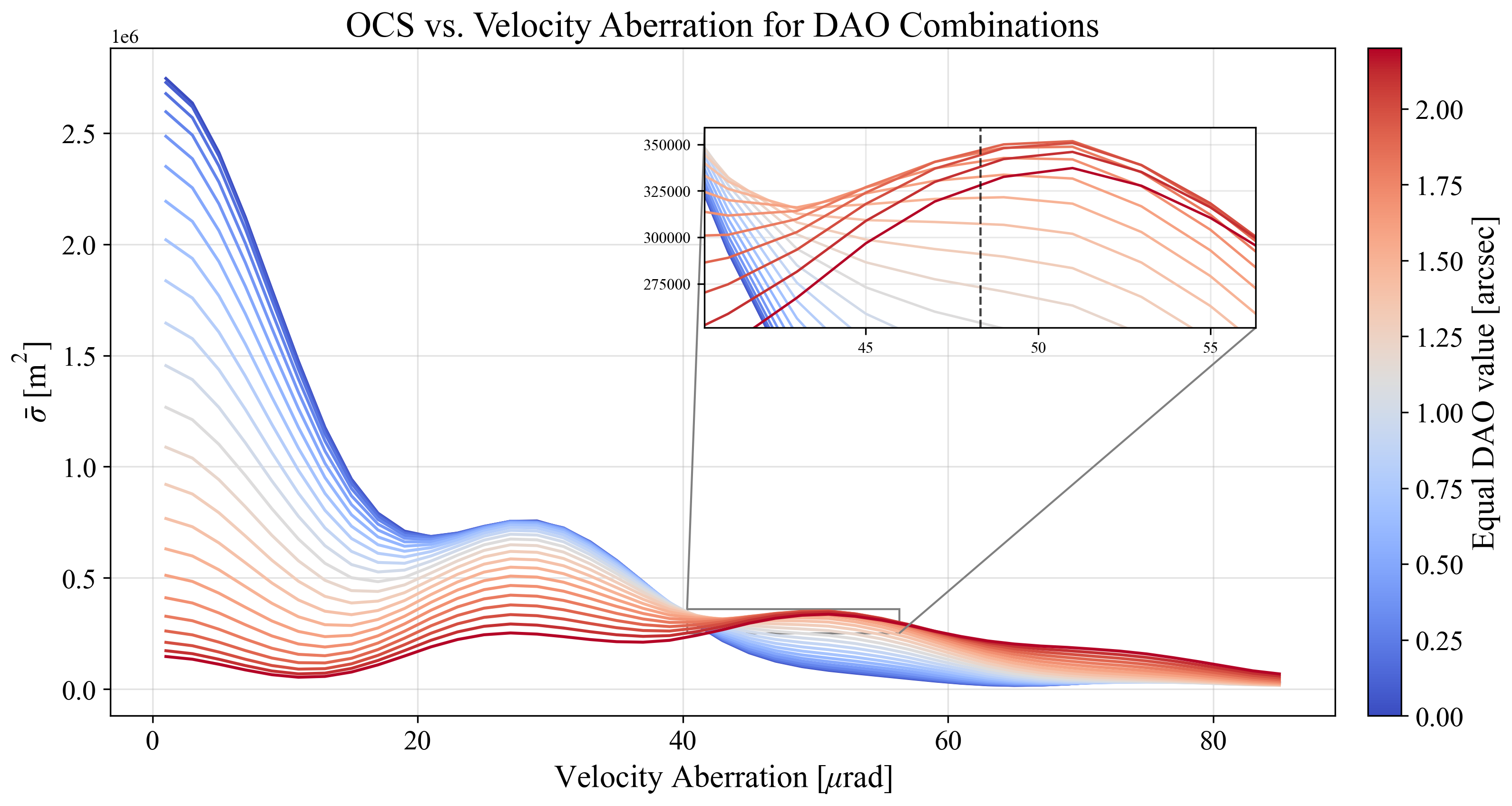}
\caption{$\bar \sigma(\mu)$ curves for CCR1s with combination of three equal DAOs with values $(0:0.1:2.3)\,\mathrm{arcsec}$.}
    \label{fig:dao_optimization_ocs_va}
\end{figure}

All curves in the $0$--$2.2\,\mathrm{arcsec}$ DAO range are shown in Fig.~\ref{fig:dao_optimization_ocs_va}, together with a zoomed view restricted to the $1.9 \,\mathrm{arcsec}$ curve around the $\mu_p$ region.

\section{Mechanical configuration: inclusion of CCR2 protrusions}
\label{sec:mechanical_configuration}

The baseline mounting concept places the CCR2s on protrusions extending outward from the spherical surface, so that the reflective tips of the $2\,\mathrm{in}$ CCRs and $1\,\mathrm{in}$ CCRs are located at the same radial distance. This solution is intended to reduce time-detection biases and ambiguities in the computation of the laser range correction and in the subsequent processing of the laser observations~\cite{otsubo2003a,Degnan2023}.

This point is particularly important for GUEST because the relative optical weight of CCR1s and CCR2s changes along the orbit. Near perigee, the total OCS is dominated by the CCR1 contribution; near apogee, the CCR2 contribution becomes more important. If the two CCR families had different effective reflection radii, the optical centre of response of the satellite would acquire an additional dependence on altitude, velocity aberration and wavelength. In other words, the laser range correction would not be determined only by the spacecraft geometry, but also by the orbit-dependent optical weighting of the two CCR populations.

Nevertheless, the advantages of this configuration with respect to alternatives without protrusions, such as LAGEOS and LARES, or with CCR1s recessed into the surface, will be fully evaluated during Phase A/B. From a dynamical perspective, the choice of the mechanical configuration affects the moment of inertia of the whole spacecraft, while, from an optical perspective, the protrusions are expected to shade only portions of the CCRs located near the outer edge of the field of view. Therefore, they are not expected to significantly affect the total spacecraft OCS, because these CCRs contribute less to the total return as a direct consequence of the incidence function $\bar k^{\mathrm{inc}}(\theta)$. 

The protruded configuration is shown in Fig.~\ref{fig:protruded_ccr2_31_config}.

The final mechanical assessment will require a shadowing analysis, together with a refined computation of the range correction. Note that the environment of the CCR mounts inherits from the positive SLR experience of LARES-2 (1-inch CCRs mounting scheme), LAGEOS, Apollo, and G2G (2-inch CCRs mounting scheme, scaled for different diameters) retroreflector arrays, where thermal loads, FFDP modifications and optical response are coupled quantities~\cite{boni_thermoptisim_2015,ciocci_moonlight2_2017,arnold_thermal_optical_2020, DellAgnello2011}.

% This step is the natural continuation of the integrated thermal-optical approach already applied to GNSS retroreflector arrays and next-generation laser retroreflectors, where thermal loads, FFDP modifications and optical response are treated as coupled quantities 

\section{Conclusions}
\label{sec:conclusions}

Equation~\eqref{eq:total_sc_ocs} approximates the total spacecraft OCS as the linear superposition of the OCS contributions from the individual CCRs. The result depends on the selected LoS, with $1 \leq j \leq N_{\mathrm{LoS}}$, where $N_{\mathrm{LoS}}$ is the total number of virtual lines of sight. The averaged quantity used to compare the different configurations is therefore
$$
\bar \sigma_{\mathrm{S/C}}(\mu)
=
\frac{1}{N_{\mathrm{LoS}}}
\sum_{j=1}^{N_{\mathrm{LoS}}}
\sigma_{\mathrm{S/C},j}(\mu).
$$

The model was first calibrated by reconstructing the LARES-2 optical response. The reconstructed OCS was found to be lower than the reference value by approximately $30\%$, which is consistent with the simplified treatment of CCR clocking adopted at this stage. The corresponding calibration factor $C_{\mathrm{LARES\text{-}2}}$ was then applied to the GUEST estimates. Within this approximation, the model is sufficiently accurate for a preliminary opto-geometrical trade-off, although it is not intended to replace the final full transfer-function computation, not applicable during the early stages of the project.

\begin{figure}[htbp]
    \centering
    \includegraphics[width=0.75\textwidth, trim=2cm 0.5cm 2cm 0.5cm, clip]{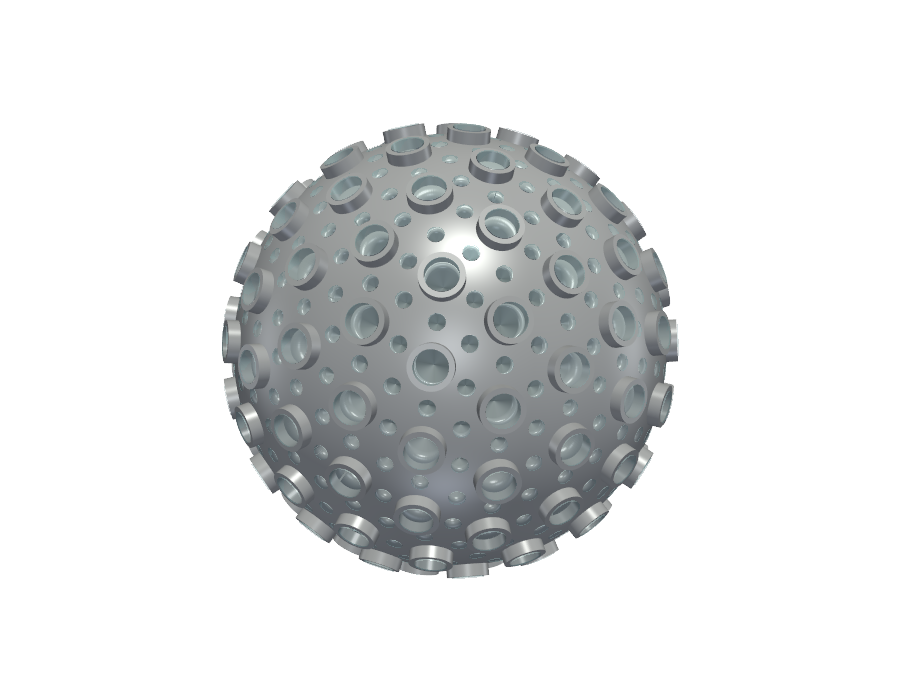}
    \caption{Artistic rendering of the spacecraft showing the distribution of CCR1s and CCR2s in the baseline $3{:}1$ configuration, with protruding CCR2s and a spacecraft radius of 35 cm.}
    \label{fig:rendering_spacecraft}
\end{figure}

The results show that a dual-CCR population is required by the GUEST orbital regime. CCR1s are more effective at perigee, where the velocity aberration is large, while CCR2s are more effective at apogee, where the velocity aberration is small and the link budget is penalized by the large slant range. The baseline $3{:}1$ CCR1/CCR2 configuration therefore represents a compromise between the two limiting observing conditions. Figure~\ref{fig:rendering_spacecraft} presents an artistic rendering of the spacecraft equipped with this configuration, featuring a radius of 35 cm and protruding CCR2s.

The sensitivity analyses lead to the following conclusions:
\begin{itemize}
    \item[\ding{226}] The infrared case improves the perigee OCS because the FFDP is broader at $\lambda=1064\,\mathrm{nm}$, but it reduces the apogee OCS because the response at small velocity aberration is less concentrated.
    \item[\ding{226}] The $8{:}1$ configuration increases the perigee OCS by increasing the relative number of CCR1s, but decreases the apogee OCS because fewer CCR2s are available.
    \item[\ding{226}] Increasing the spacecraft radius from $35\,\mathrm{cm}$ to $43\,\mathrm{cm}$ is the only variation considered here that increases the OCS at both perigee and apogee.
    \item[\ding{226}] Introducing equal non-null DAOs in the CCR1 population can optimize the response around $\mu_p$, but the improvement is localized for a specific VA.
    
    \item[\ding{226}] Placing CCR2s on protrusions is mechanically more complex, but it reduces differential timing ambiguities between the two CCR families and simplifies the interpretation of the laser range correction.
\end{itemize}

Overall, the baseline configuration provides a balanced solution for perigee visibility, apogee visibility, isotropy and mechanical feasibility. The larger-satellite option gives the strongest global OCS improvement, while the infrared, $8{:}1$ and non-null-DAO configurations are more targeted optimizations of the perigee response.

The present study should therefore be considered as a reduced-order design and optimization step. The final Phase A/B analysis will require explicit CCR clocking, protrusion shadowing, thermal gradients, manufacturing tolerances, polarization effects, and a complete computation of the range-correction transfer function. These refinements follow the same methodological line developed for LAGEOS, LARES, GNSS retroreflector arrays and next-generation laser retroreflector payloads~\cite{arnold_transfer_functions,arnold_lares_transfer_2015,arnold_thermal_optical_2020,boni_thermoptisim_2015,ciocci_moonlight2_2017}.

\section*{Acknowledgments}
%S.D.A. wishes to thank ASI for the support through ASI-INFN Agreement n. 2019-15-HH.0 (Joint Lab), ASI-INFN Agreement n. 2023-59-HH.0 (LaGREx), and through  ASI‐UNIBO-INFN Agreement n. 2022‐8‐HH.0 (Hera mission of ESA) on the studies on lunar laser retroreflector and LLR presented here. 
S. Dell'Agnello, L. Porcelli contributed upon initial request of the GUEST PI in the framework of their research activities as users and members of the ILRS.
G. Bargiacchi, N. Burzill\`a, acknowledge the support from the Italian Space Agency (ASI) through the research Agreement ASI-INFN n. 2019-15-HH.0 until April 1, 2026.
D. Vetrano, F. Del Ferraro Iacobelli acknowledge the support from ASI through the research Agreement ASI n. 2022‐8‐HH.0 (science studies for ESA's Hera mission).
R. Campagnola acknowledges the support from the European Space Agency (ESA) through the contract ESA-INFN Contract n. 4000133721/21/NL/CR.
IFAE is partially funded by the CERCA program of the Generalitat de Catalunya.
This publication is part of the R\&D\&i project PID2023-146686NB-C31 funded by MICIU/AEI/10.13039/501100011033/ and by ERDF/EU.
D. Blas acknowledges financial support from the Spanish Ministry of Science and Innovation (MICINN) through the Spanish State Research Agency, under Severo Ochoa Centres of Excellence Programme 2025-2029 (CEX2024001442-S).
This work is supported by ERC grant (GravNet, ERC-2024-SyG 101167211, DOI: 10.3030/101167211). Funded by the European Union. Views and opinions expressed are, however those of the author(s) only and do not necessarily reflect those of the European Union or the European Research Council Executive Agency. Neither the European Union nor the granting authority can be held responsible for them. 
D. Blas acknowledges the support from the European Research Area (ERA) via the UNDARK project of the Widening participation and spreading excellence programme (project number 101159929).

\bibliographystyle{apsrev4-2}
\bibliography{articles}

@article{Degnan2023,
  author       = {Degnan, John J.},
  title        = {A Tutorial on Retroreflectors and Arrays Used in Satellite and Lunar Laser Ranging},
  journal = {Photonics},
  volume       = {10},
  number       = {11},
  pages        = {1215},
  year         = {2023},
  doi          = {10.3390/photonics10111215},
  url          = {https://www.mdpi.com/2304-6732/10/11/1215}
}

@inproceedings{arnold_transfer_functions,
  author    = {David A. Arnold},
  title     = {Retroreflector Array Transfer Functions},
  booktitle = {Proceedings of the 13th International Workshop on Laser Ranging},
  location  = {Washington, DC},
  note      = {October 7--11},
  year      = {2002}
}

@report{arnold_lares_transfer_2015,
  author      = {David A. Arnold},
  title       = {Transfer Function of the LARES Retroreflector Array},
  institution = {ILRS},
  type        = {Technical Report},
  note        = {February},
  year        = {2015}
}

@article{arnold_thermal_optical_2020,
  author       = {David A. Arnold},
  title        = {Thermal-optical design of a geodetic satellite for one millimeter accuracy},
  journal = {Adv. Space Res.},
  volume       = {65},
  pages        = {2276--2289},
  year         = {2020}
}

@article{boni_thermoptisim_2015,
  author       = {Alessandro Boni and Claudio Cantone and Simone Dell'Agnello and Emanuele Ciocci and Stefania Contessa and Giovanni O. Delle Monache and Cristina Lops and Marco Maiello and Manuele Martini and Claudia Mondaini and Giuseppe Patrizi and Luca Porcelli and Lorenzo Salvatori and Marco Tibuzzi and Pierguido Tuscano and Roberto Vittori and Giuseppe Bianco},
  title        = {Integrated thermal-optical simulations of a GNSS Retroreflector Array},
  journal = {Adv. Space Res.},
  volume       = {56},
  pages        = {85--91},
  year         = {2015}
}

@article{ciocci_moonlight2_2017,
  author       = {Emanuele Ciocci and Manuele Martini and Stefania Contessa and Luca Porcelli and Marco Mastrofini and Douglas Currie and Giovanni O. Delle Monache and Simone Dell'Agnello},
  title        = {Performance analysis of next-generation lunar laser retroreflectors},
  journal = {Adv. Space Res.},
  volume       = {60},
  pages        = {1300--1306},
  year         = {2017}
}

@unpublished{GUEST_white_paper,
  author = {Diego Blas and Francisco J. Atapuerca and Giada Bargiacchi and Massimo Bassan and Bruno Bertrand and Mathis Blossfeld and Adrien Bourgoin and Nicolò Burzillà and Alessandro Bursi and Alfonso Caldiero and Roberto Campagnola and Ana Caramete and Laurentiu Caramete and Joan Manel Casalta Escuer and Julien Chabé and Gabriel Chiritoi and Marco Cinelli and Florin Ioan Constantin and Emiliano Cordelli and Neil Cornish and Clement Courde and Simone Dell'Agnello and Alessandro Di Marco and Andrea Di Mira and Sebastian Ellis and Malcolm Fairbairn and Ariadna Farrés and Joshua W. Foster and Marta Goli and Michael H"afner and Daniel Hampf and Aurélien Hees and Clemens Heese and Maria Isfan and Justin Janquart and Alexander C. Jenkins and Sébastien Le Maistre and David Lucchesi and Marco Lucente and Jorge Martìn Camalich and Rosa Martìnez Rubiella and Josep Mas Masdemont and Toshimichi Otsubo and Crist'obal Padilla and Alice Paun and Alice Perego and Roberto Peron and Florentina Pislan and Florin Adrian Popescu and Luca Porcelli and Nanda Rea and Rafael Rebolo and Marco Reyes and Ignasi Ribas and José C. Rodrìguez and Pascal Rosenblatt and Albert Roura and Soumen Roy and Mariano Sanchez Nogales and Francesco Santoli and Feliciana Sapio and Anja Schlicht and Ulrich Schreiber and Angela Serrano and Daniel Serrano Lombillo and Alberto Sesana and Carlos Sopuerta and Krzysztof So'snica and Matteo Spinelli and Nicola Tamanini and Dario Vetrano and Massimo Visco and Hanxi Wang and Xiao Xue and Miguel Zumalac'arregui},
  title  = {GUEST: Gravitational Universe Exploration with Satellite Tracking: A passive satellite laser ranging for the dark gravitational universe},
  note   = {arXiv:submit/7841867 [astro-ph.CO] 19 Jul 2026}
}

@article{Blas2022,
  author       = {D. Blas and A. Jenkins},
  title        = {Bridging the \ensuremath{\mu}Hz Gap in the Gravitational-Wave Landscape with Binary Resonances},
  journal = {Phys. Rev. Lett.},
  volume       = {128},
  pages        = {101103},
  year         = {2022},
  note         = {arXiv:2107.04601}
}

@article{Blas2022a,
  author       = {D. Blas and A. C. Jenkins},
  title        = {Detecting stochastic gravitational waves with binary resonance},
  journal = {Phys. Rev. D},
  volume       = {105},
  pages        = {064021},
  year         = {2022},
  note         = {arXiv:2107.04063}
}

@misc{fosterBlas2025,
      title={Discovering $\mu$Hz gravitational waves and ultra-light dark matter with binary resonances}, 
      author={Joshua W. Foster and Diego Blas and Adrien Bourgoin and Aurelien Hees and Míriam Herrero-Valea and Alexander C. Jenkins and Xiao Xue},
      year={2025},
      eprint={2504.15334},
      archivePrefix={arXiv},
      primaryClass={astro-ph.CO},
      url={https://arxiv.org/abs/2504.15334}, 
}

@unpublished{Foster2025a,
  author = {J. W. Foster and others},
  title  = {Prospects for gravitational wave and ultra-light dark matter detection with binary resonances beyond the secular approximation},
  note   = {submitted to Phys. Rev. D (2025), arXiv:2504.16988 [gr-qc]}
}

@unpublished{Foster2025b,
  author = {J. W. Foster and others},
  title  = {Discovering {$\mu$Hz} gravitational waves and ultra-light dark matter with binary resonances},
  note   = {submitted to Phys. Rev. Lett. (2025), arXiv:2504.15334 [astro-ph.CO]}
}

@article{Hui2013,
  author       = {L. Hui and others},
  title        = {Binary systems as resonance detectors},
  journal = {Phys. Rev. D},
  volume       = {87},
  pages        = {084009},
  year         = {2013},
  note         = {arXiv:1212.2623}
}

@article{ciufolini_lares2,
  author       = {Ignazio Ciufolini and Antonio Paolozzi and Erricos C. Pavlis and John C. Ries and Richard Matzner and Claudio Paris and Emiliano Ortore and Vahe Gurzadyan and Roger Penrose},
  title        = {The LARES 2 satellite, general relativity and fundamental physics},
  journal = {Eur. Phys. J. C},
  volume       = {83},
  pages        = {87},
  year         = {2023}
}

@article{pirrotta2025,
title = {LARES 2 mission: The consolidation of Italian heritage in laser ranged satellites},
journal = {Acta Astronautica},
volume = {235},
pages = {787-797},
year = {2025},
issn = {0094-5765},
doi = {https://doi.org/10.1016/j.actaastro.2025.06.006},
url = {https://www.sciencedirect.com/science/article/pii/S0094576525003546},
author = {Simone Pirrotta and Giuseppe Bianco and Ignazio Ciufolini and Roberto Bertacin and Alessandro Bursi and Simone Dell’Agnello and Antonio Paolozzi and Rocco C. Pellegrini and Adriano Pepato and Matteo Spinelli},
}

@online{ILRS_LAGEOS,
  author       = {{International Laser Ranging Service}},
  title        = {LAGEOS-1, -2},
  organization = {International Laser Ranging Service},
  url          = {https://ilrs.gsfc.nasa.gov/missions/satellite_missions/current_missions/lag1_general.html},
  urldate      = {2026-06-19}
}

@online{ILRS_LARES,
  author       = {{International Laser Ranging Service}},
  title        = {LARES -- LAser RElativity Satellite},
  organization = {International Laser Ranging Service},
  url          = {https://ilrs.gsfc.nasa.gov/missions/satellite_missions/current_missions/lrs2_general.html},
  urldate      = {2026-06-19}
}

@article{Otsubo2011,
author       = {Otsubo, Toshimichi and Kunimori, Hiroo and Noda, Hirotomo and Hanada, Hideo and Araki, Hiroshi and Katayama, Masato},
title        = {Asymmetric dihedral angle offsets for large-size lunar laser ranging retroreflectors},
journal = {Earth, Planets and Space},
volume       = {63},
number       = {8},
pages        = {e13--e16},
year         = {2011},
doi          = {10.5047/eps.2011.11.001},
url          = {https://doi.org/10.5047/eps.2011.11.001},
issn         = {1880-5981}
}

@article{Williams_2023,
author = {Williams, James G. and Porcelli, Luca and Dell’Agnello, Simone and Mauro, Lorenza and Muccino, Marco and Currie, Douglas G. and Wellnitz, Dennis and Wu, Chensheng and Boggs, Dale H. and Johnson, Nathan H.},
title = {Lunar Laser Ranging Retroreflectors: Velocity Aberration and Diffraction Pattern},
journal = {The Planetary Science Journal},
volume = {4},
number = {5},
pages = {89},
doi = {10.3847/PSJ/acbeab},
url = {https://doi.org/10.3847/PSJ/acbeab},
year = {2023},
month = {may},
publisher = {The American Astronomical Society},
}

@article{otsubo2003a,
author = {Otsubo, Toshimichi and Appleby, Graham M.},
title = {System-dependent center-of-mass correction for spherical geodetic satellites},
journal = {Journal of Geophysical Research: Solid Earth},
volume = {108},
number = {B4},
pages = {},
doi = {https://doi.org/10.1029/2002JB002209},
url = {https://agupubs.onlinelibrary.wiley.com/doi/abs/10.1029/2002JB002209},
eprint = {https://agupubs.onlinelibrary.wiley.com/doi/pdf/10.1029/2002JB002209},
year = {2003}
}

@article{thomson_solutions,
  title = {Structure and dynamics of spherical crystals characterized for the Thomson problem},
  author = {Wales, David J. and Ulker, Sidika},
  journal = {Phys. Rev. B},
  volume = {74},
  issue = {21},
  pages = {212101},
  numpages = {4},
  year = {2006},
  month = {Dec},
  publisher = {American Physical Society},
  doi = {10.1103/PhysRevB.74.212101},
  url = {https://link.aps.org/doi/10.1103/PhysRevB.74.212101}
}

@article{points_on_sphere,
  author  = {Saff, E. B. and Kuijlaars, A. B. J.},
  title   = {Distributing Many Points on a Sphere},
  journal = {The Mathematical Intelligencer},
  volume  = {19},
  number  = {1},
  pages   = {5--11},
  year    = {1997},
  doi     = {10.1007/BF03024331},
  url     = {https://doi.org/10.1007/BF03024331},
  issn    = {1866-7414}
}

@article{fibonacci_sphere,
author = {Keinert, Benjamin and Innmann, Matthias and S\"{a}nger, Michael and Stamminger, Marc},
title = {Spherical fibonacci mapping},
year = {2015},
issue_date = {November 2015},
publisher = {Association for Computing Machinery},
address = {New York, NY, USA},
volume = {34},
number = {6},
issn = {0730-0301},
url = {https://doi.org/10.1145/2816795.2818131},
doi = {10.1145/2816795.2818131},
journal = {ACM Trans. Graph.},
month = nov,
articleno = {193},
numpages = {7}
}

@book{Goodman2005,
  author    = {Goodman, Joseph W.},
  title     = {Introduction to Fourier Optics},
  edition   = {3},
  publisher = {Roberts and Company Publishers},
  address   = {Englewood, Colorado},
  year      = {2005}
}

@unpublished{LISA2024,
  author = {{LISA Collaboration}},
  title  = {LISA Definition Study Report},
  note   = {arXiv:2402.07571 [astro-ph.CO]},
  year   = {2024}
}

@article{Delva2023,
  author       = {P. Delva and others},
  title        = {GENESIS: co-location of geodetic techniques in space},
  journal = {Earth Planets Space},
  volume       = {75},
  pages        = {5},
  year         = {2023}
}

@article{NANOGrav2023,
  author       = {{G. Agazie et al. (NANOGrav)}},
  title        = {Evidence for a gravitational-wave background},
  journal = {Astrophys. J. Lett.},
  volume       = {951},
  pages        = {L8},
  year         = {2023}
}

@article{LIGO2016,
  author       = {B. P. Abbott and others},
  title        = {Observation of gravitational waves from a binary black hole merger},
  journal = {Phys. Rev. Lett.},
  volume       = {116},
  pages        = {061102},
  year         = {2016}
}

@article{DellAgnello2011,
  author  = {Dell'Agnello, S. and Delle Monache, G. and Currie, D. and
             Vittori, R. and Cantone, C. and Garattini, M. and Boni, A. and
             Martini, M. and Lops, C. and Intaglietta, N. and Tauraso, R. and
             Arnold, D. and Pearlman, M. and Bianco, G. and Zerbini, S. and
             Maiello, M. and Berardi, S. and Porcelli, L.},
  title   = {Creation of the new industry-standard space test of laser
             retroreflectors for the {GNSS} and {LAGEOS}},
  journal = {Advances in Space Research},
  volume  = {47},
  year    = {2011},
  pages   = {822--842}
}

% \newpage

\end{document}